\documentclass[11pt]{article}
\usepackage[utf8]{inputenc}
\usepackage{amsfonts}
\usepackage{amsmath}
\usepackage{amssymb}
\usepackage{indentfirst}
\usepackage{graphicx}
\usepackage{hyperref}
\usepackage{cite}
\usepackage{color}
\usepackage{xcolor}

\numberwithin{equation}{section}
\allowdisplaybreaks
\makeatletter

\begin{document}

\begin{titlepage}
\vskip 4cm

\begin{center}
\textbf{\LARGE{Three-dimensional non-relativistic extended supergravity with cosmological constant}}
\par\end{center}{\LARGE \par}

\begin{center}
	\vspace{1cm}
	\textbf{Patrick Concha}$^{\ast}$,
    \textbf{Lucrezia Ravera}$^{\ddag,\dag}$,
	\textbf{Evelyn Rodríguez}$^{\star}$,
	\small
	\\[5mm]
	$^{\ast}$\textit{Departamento de Matemática y Física Aplicadas, }\\
	\textit{ Universidad Católica de la Santísima Concepción, }\\
\textit{ Alonso de Ribera 2850, Concepción, Chile.}
	\\[3mm]
    $^{\ddag}$\textit{DISAT, Politecnico di Torino, }\\
	\textit{Corso Duca degli Abruzzi 24, 10129 Torino, Italy.}
	\\[3mm]
	$^{\dag}$\textit{INFN, Sezione di Torino, }\\
	\textit{Via P. Giuria 1, 10125 Torino, Italy.}
	\\[3mm]
	$^{\star}$\textit{Departamento de Ciencias, Facultad de Artes Liberales,} \\
	\textit{Universidad Adolfo Ibáñez, Viña del Mar, Chile.} \\[5mm]
	\footnotesize
	\texttt{patrick.concha@ucsc.cl},
    \texttt{lucrezia.ravera@polito.it},
	\texttt{evelyn.rodriguez@edu.uai.cl},
	\par\end{center}
\vskip 20pt
\centerline{{\bf Abstract}}
\medskip
\noindent

In this paper, we present two novel non-relativistic superalgebras which correspond to supersymmetric extensions of the enlarged extended Bargmann algebra.  The three-dimensional non-relativistic Chern-Simons supergravity actions invariant under the aforementioned superalgebras are constructed. The new non-relativistic superalgebras allow to accommodate a cosmological constant in a non-relativistic supergravity theory. Interestingly, we show that one of the non-relativistic supergravity theories presented here leads to the  recently introduced  Maxwellian exotic Bargmann supergravity when the flat limit $\ell \rightarrow\infty$ is considered. Besides, we show that both descriptions can be written in terms of a supersymmetric extension of the Nappi-Witten algebra or the extended Newton-Hooke superalgebra.

\end{titlepage}\newpage {\baselineskip=12pt \tableofcontents{}}

\section{Introduction}

Non-relativistic (NR) supergravity theories have recently been studied and their construction has only been approached in three spacetime dimensions \cite{Andringa:2013mma,Bergshoeff:2015ija,Bergshoeff:2016lwr,Ozdemir:2019orp,Ozdemir:2019tby,Concha:2019mxx}. In this way, to formulate a NR (super)gravity
theory invariant under a certain NR (super)symmetry remains as a challenging
but at the same time interesting problem. At the bosonic level, the NR theories have been of particular interest due to their relation to condensed matter systems \cite{Son:2008ye,Balasubramanian:2008dm,Kachru:2008yh,Bagchi:2009my,Bagchi:2009pe,Christensen:2013lma,Christensen:2013rfa,Hartong:2014oma,Hartong:2014pma,Hartong:2015wxa,Taylor:2015glc} and NR effective field theories \cite{Hoyos:2011ez,Son:2013rqa,Abanov:2014ula,Geracie:2015dea,Gromov:2015fda}. In presence of supersymmetry, moving toward this kind of theories can be useful, for instance, to approach supersymmetric field
theories on non-relativistic curved backgrounds via localization \cite{Festuccia:2011ws,Pestun:2007rz,Marino:2012zq}.   Also from the purely theoretical point of view, NR supergravity is an
attractive topic that deserves to be developed. The construction of NR
(super)algebras and the formulation of NR (super)gravity theories are of
special interest for the present work. As it happens in NR gravity theories, the
suitable construction of a NR supergravity action has been shown to require the
introduction of extra bosonic and fermionic generators. In particular,
working in three spacetime dimensions, the appropriate construction of a
Chern-Simons (CS) action is possible due to the presence of additional generators
which allow to have a non-degenerate invariant tensor. The non-degeneracy of the invariant tensor is a fundamental
ingredient to get a well-defined CS (super)gravity action \cite{Witten:1988hc,Achucarro:1987vz,Zanelli:2005sa}.

Recently, the NR version of a three-dimensional CS gravity theory invariant
under a particular enlargement of the extended Bargmann algebra \cite{LevyLeblond,Grigore:1993fz,Bose:1994sj,Duval:2000xr,Jackiw:2000tz,Horvathy:2002vt,Papageorgiou:2009zc} was presented
in \cite{Concha:2019lhn}. Such algebra, called as enlarged extended Bargmann
(EEB) algebra by the authors, was obtained by considering the NR contraction
of the [AdS-Lorentz]$ \, \oplus \, \mathfrak{u}_1 \oplus \mathfrak{u}_1 \oplus \mathfrak{u}_1$ algebra. Such novel NR symmetry offers us an alternative way to accomodate a cosmological constant to a three-dimensional NR gravity theory different from those studied in \cite{Aldrovandi:1998im,Gibbons:2003rv,Brugues:2006yd,Alvarez:2007fw,Duval:2011mi,Papageorgiou:2010ud,Hartong:2016yrf,Duval:2016tzi,Concha:2019dqs}. Interestingly, as was shown in \cite{Concha:2019lhn}, the Maxwellian exotic Bargmann (MEB) algebra recently introduced in \cite{Aviles:2018jzw} can be obtained from the EEB
gravity theory when the flat limit $\ell \rightarrow \infty $ is considered (where $\ell$ is the length parameter related to the cosmological constant through $\Lambda \propto \pm \frac{1}{\ell^2}$; the limit $\ell \rightarrow \infty $ corresponds to $\Lambda \rightarrow 0 $).

The AdS-Lorentz (AdS-$\mathcal{L}$) (super)symmetry has been introduced in \cite{Soroka:2004fj,Soroka:2006aj,Gomis:2009dm} and has been useful in three and higher dimensions
in diverse (super)gravity contexts. On the gravity side, this symmetry and
its generalizations were used to recover (pure) Lovelock gravities from CS
and Born-Infeld theories in different odd an even dimensions, respectively
\cite{Concha:2016kdz,Concha:2016tms,Concha:2017nca}. Subsequently, in the case of a
three-dimensional gravity theory invariant under the AdS-$\mathcal{L}$ algebra, it has been shown that the asymptotic symmetry of conserved charges at null infinity is given by a
semi-simple enlargement of $\mathfrak{bms}_{3}$\cite{Concha:2018jjj}. On the other
hand, the supersymmetric extension of the AdS-$\mathcal{L}$ algebra allowed to introduce a generalized supersymmetric cosmological constant term in a four-dimensional supergravity theory \cite{Concha:2015tla,Ipinza:2016con,Banaudi:2018zmh,Penafiel:2018vpe,Concha:2018ywv}. In three spacetime dimensions, the construction of the $\mathcal{N}$-extended AdS-$\mathcal{L}$ supergravity theory was presented in \cite{Concha:2018jxx,Concha:2019icz}.

As pointed out in \cite{Concha:2019lhn}, an open problem was to extend their results at the supersymmetric level. Very recently, a NR version of the Maxwell CS supergravity theory was presented in \cite{Concha:2019mxx} while the construction of a possible EEB supergravity theory remained unexplored till now.
In this work, we approach the aforesaid problem, and present two different supersymmetric extensions of the EEB algebra by expanding different relativistic $\mathcal{N}=2$ AdS-$\mathcal{L}$ superalgebras. In particular, the Lie algebra expansion method \cite{Hatsuda:2001pp,deAzcarraga:2002xi,deAzcarraga:2007et,Izaurieta:2006zz} has resulted to be a powerful tool in NR symmetries \cite{Concha:2019lhn,Bergshoeff:2019ctr,deAzcarraga:2019mdn,Penafiel:2019czp,Romano:2019ulw,Harmark:2019upf,Gomis:2019nih,Bergshoeff:2020fiz,Kasikci:2020qsj,Concha:2020sjt,Fontanella:2020eje,Concha:2020ebl}. Here, we show that the semigroup expansion method \cite{Izaurieta:2006zz} allows us to obtain not only new NR superalgebras but also provides with
the non-vanishing components of the invariant tensor allowing to construct the corresponding NR CS supergravity actions. We show that, although both supersymmetric descriptions allow us to introduce a cosmological constant to a NR supergravity action, only one supersymmetric extension contains a well-defined vanishing cosmological constant limit.

The paper is organized as follows: In Section 2, a brief review of the EEB
algebra and the corresponding formulation of the three-dimensional CS
gravity action are presented. Sections 3 and 4 contain the main results of the
work. First, a supersymmetric extension of the EEB is presented. Then, a NR
CS supergravity action based on the EEB superalgebra is constructed. The
supersymmetry transformation laws are also provided. In Section 4, we provide with an alternative supersymmetric extension of the EEB algebra which we call non-standard EEB superalgebra. The construction of the CS supergravity action is also studied. Section 5 concludes our work with discussions and some possible future approaches.

\section{Enlarged Extended Bargmann gravity}

In this section, we briefly review the enlarged extended Bargmann algebra
considered in \cite{Concha:2019lhn} and the associated CS gravity
theory constructed in the same paper in three spacetime dimensions.
The particular enlargment of the Extended Bargmann algebra presented in \cite%
{Concha:2019lhn} was denoted as EEB algebra and its non-trivial commutation relations read
\begin{eqnarray}
\left[ \tilde{G}_{a},\tilde{P}_{b}\right] &=&-\epsilon _{ab}\tilde{M}%
\,,\qquad \left[ \tilde{G}_{a},\tilde{Z}_{b}\right] = -\epsilon _{ab}\tilde{T%
}\,, \qquad \left[ \tilde{H}, \tilde{Z}_a \right] = \frac{1}{\ell^2}
\epsilon_{ab} \tilde{P}_b \, ,  \notag \\
\left[ \tilde{H},\tilde{G}_{a}\right] &=&\epsilon _{ab}\tilde{P}%
_{b}\,,\qquad \quad \ \left[ \tilde{J},\tilde{Z}_{a}\right] =\epsilon _{ab}%
\tilde{Z}_{b}\,, \ \ \ \ \quad \left[ \tilde{Z}_a , \tilde{Z}_b \right] = -
\frac{1}{\ell^2} \epsilon_{ab} \tilde{T} \,,  \notag \\
\left[ \tilde{J},\tilde{P}_{a}\right] &=&\epsilon _{ab}\tilde{P}%
_{b}\,,\qquad \quad \left[ \tilde{H},\tilde{P}_{a}\right] =\epsilon _{ab}%
\tilde{Z}_{b}\,, \ \qquad \left[ \tilde{P}_a , \tilde{Z}_b \right] = - \frac{%
1}{\ell^2} \epsilon_{ab} \tilde{M} \,,  \notag  \\
\left[ \tilde{J},\tilde{G}_{a}\right] &=&\epsilon _{ab}\tilde{G}%
_{b}\,,\qquad \ \ \left[ \tilde{P}_{a},\tilde{P}_{b}\right] =-\epsilon _{ab}%
\tilde{T}\,, \qquad \ \left[ \tilde{Z},\tilde{P}_{a}\right] = \frac{1}{\ell^2%
}\epsilon _{ab}\tilde{P}_{b} \,,  \notag \\
\left[ \tilde{G}_{a},\tilde{G}_{b}\right] &=&-\epsilon _{ab}\tilde{S}%
\,,\qquad \ \left[ \tilde{Z},\tilde{G}_{a}\right] =\epsilon _{ab}\tilde{Z}%
_{b} \,, \qquad \ \ \ \left[ \tilde{Z},\tilde{Z}_{a}\right] =\frac{1}{\ell^2}%
\epsilon _{ab}\tilde{Z}_{b}\,,  \label{EEB1}
\end{eqnarray}
where $a=1,2$, $\epsilon_{ab}=\epsilon_{0ab}$ and $\epsilon^{ab}=\epsilon^{0ab}$, and where $\ell$ is a length parameter. Such algebra turns out to be the NR version of the relativistic [AdS-$\mathcal{L}$]$\, \oplus  \, \mathfrak{u}(1)\oplus\mathfrak{u}(1)\oplus\mathfrak{u}(1)$ algebra. The AdS-$\mathcal{L}$ algebra reads
\begin{eqnarray}
\left[ J_{A},J_{B}\right] &=&\epsilon _{ABC}J^{C}\,, \qquad \qquad \, \left[ J_{A},P_{B}\right] =\epsilon _{ABC}P^{C}\,, \notag \\
\left[ J_{A},Z_{B}\right] &=&\epsilon _{ABC}Z^{C}\,, \qquad \qquad \left[ P_{A},P_{B}\right] =\epsilon _{ABC}Z^{C}\,, \notag \\
\left[ Z_{A},Z_{B}\right] &=&\frac{1}{\ell ^{2}}\epsilon _{ABC}Z^{C}\,, \qquad \ \ \, \left[ P_{A},Z_{B}\right] =\frac{1}{\ell ^{2}}\epsilon _{ABC}P^{C}\,,  \label{AdSL}
\end{eqnarray}%
with $A=0,1,2$, where $J_A$ are the Lorentz generators, $P_A$ the spacetime translation generators, and $Z_A$ are extra Abelian charges. The aforementioned $U(1)$-enlargement (given in terms of the $U(1)$ generators $Y_1$, $Y_2$, and $Y_3$) of the relativistic AdS-$\mathcal{L}$ symmetry is necessary to assure a finite and non-degenerate invariant tensor allowing to construct a well-defined NR CS gravity theory in three spacetime dimensions. As it was shown in \cite{Concha:2019lhn}, the EEB algebra \eqref{EEB1} appears as a contraction of the relativistic algebra \eqref{AdSL}. Indeed, the NR algebra is obtained considering the limit $\xi\rightarrow\infty$ in the following redefinition of the [AdS-$\mathcal{L}$]$ \, \oplus \, \mathfrak{u}(1)\oplus\mathfrak{u}(1)\oplus\mathfrak{u}(1)$ generators:
\begin{eqnarray}
J_0&=&\frac{\tilde{J}}{2}+\xi^{2}\tilde{S}\,,\qquad \quad J_{a}=\xi\tilde{G}_{a}\,, \qquad \quad Y_2=\frac{\tilde{J}}{2}-\xi^{2}\tilde{S}\,, \notag \\
P_0&=&\frac{\tilde{H}}{2\xi}+\xi\tilde{M}\,,\qquad \ \ P_{a}=\tilde{P}_{a}\,, \qquad \quad \ \, Y_1=\frac{\tilde{H}}{2\xi}-\xi\tilde{M}\,, \notag \\
Z_0&=&\frac{\tilde{Z}}{2\xi^{2}}+\tilde{T}\,,\qquad \quad \, Z_{a}=\frac{\tilde{Z}_{a}}{\xi}\,, \qquad \quad \, Y_3=\frac{\tilde{Z}}{2\xi^{2}}-\tilde{T}\,, 
\end{eqnarray}
together with the rescaling of the lenght parameter $\ell\rightarrow\xi\ell$. Furthermore, as in the case of its relativistic counterpart, a vanishing cosmological constant limit $\ell \rightarrow \infty$ can be applied to the EEB algebra \eqref{EEB1}, leading to the Maxwellian Extended Bargmann (MEB) algebra of \cite{Aviles:2018jzw}. 

The EEB algebra admits the following non-vanishing components of the invariant tensor:
\begin{eqnarray}
\left\langle \tilde{G}_a \tilde{G}_b \right\rangle &=& \tilde{\alpha}_0
\delta_{ab} \,, \qquad \quad \ \ \, \left\langle \tilde{G}_a \tilde{P}_b \right\rangle = \tilde{\alpha}_1
\delta_{ab} \,, \notag \\
\left\langle \tilde{G}_a \tilde{Z}_b \right\rangle &=& \tilde{\alpha}_2
\delta_{ab} \,, \qquad \qquad \left\langle \tilde{P}_a \tilde{P}_b \right\rangle=\tilde{\alpha}_2
\delta_{ab} \,, 
\notag \\
\left\langle \tilde{J} \tilde{S} \right\rangle &=& -\tilde{\alpha}_0 \,, \qquad \qquad \ \ \,
\left\langle \tilde{J} \tilde{M} \right\rangle = -\tilde{\alpha}_1 \,, \notag \\
\left\langle \tilde{H} \tilde{S} \right\rangle &=&-\tilde{\alpha}_1  \,,  \qquad \qquad \ \ \ \,
\left\langle \tilde{J} \tilde{T} \right\rangle = -\tilde{\alpha}_2 \,, \notag \\
\left\langle \tilde{H} \tilde{M} \right\rangle \, &=&  -\tilde{\alpha}_2 \,, \qquad \qquad \ \ \ \, \left\langle \tilde{S}  \tilde{Z} \right\rangle= -\tilde{\alpha}_2 \,,  \notag \\
\left\langle \tilde{Z}_a \tilde{Z}_b \right\rangle &=& \frac{\tilde{\alpha}_2%
}{\ell^2} \delta_{ab} \,, \qquad \quad \ \
\left\langle \tilde{Z}_a \tilde{P}_b \right\rangle = \frac{\tilde{\alpha}_1%
}{\ell^2} \delta_{ab} \,,  \notag \\
\left\langle \tilde{Z} \tilde{M} \right\rangle &=& - \frac{\tilde{\alpha}_1}{%
\ell^2} \,, \qquad \qquad \ \ \left\langle \tilde{T} \tilde{H} \right\rangle = - \frac{\tilde{\alpha}_1}{%
\ell^2} \,,  \notag \\
\left\langle \tilde{Z} \tilde{T} \right\rangle &=& - \frac{\tilde{\alpha}_2}{%
\ell^2} \,,  \label{invt1}
\end{eqnarray}
where $\tilde{\alpha}_{0}$, $\tilde{\alpha}_{1}$, and $\tilde{\alpha}_{2}$ are arbitrary constants which are related to the relativistic parameters as
\begin{equation}
\alpha_0=\tilde{\alpha}_{0}\xi^{2}\,,\qquad\alpha_1=\tilde{\alpha}_{1}\xi\,,\qquad\alpha_{2}=\tilde{\alpha}_{2}\,.
\end{equation}
Such rescaling of the coupling constants reproduces the most general NR non-degenerate invariant tensor in the limit $\xi\rightarrow\infty$. Let us note that the non-degeneracy of the invariant tensor is related to the Physical requirement that the CS action involves a kinematical term for each field and the equation of motions imply that all curvatures vanish. On the other hand, one can see that the limit $\ell \rightarrow \infty$ applied to the components of the
invariant tensor written in (\ref{invt1}) leads to the invariant tensor of
the MEB algebra \cite{Aviles:2018jzw}.

The CS form for a connection $A$ constructed with the invariant non-degenerate bilinear form
defines an action for the relativistic gauge theory based on the symmetry under
consideration as
\begin{equation}
I_{\text{CS}}=\int \langle A\wedge dA+\frac{2}{3}A\wedge A\wedge A\rangle
=\int \langle A\wedge dA+\frac{1}{3}A\wedge \left[ A,A\right] \rangle \,.
\label{genCS}
\end{equation}%
In terms of the NR generators and fields of the EEB algebra, the gauge connection one-form of \cite{Concha:2019lhn}, $A=A^{A}T_{A}$,
where $T_{A}=\{\tilde{J},\tilde{G}_{a},\tilde{H},\tilde{P}_{a},%
\tilde{Z},\tilde{Z}_{a},\tilde{M},\tilde{S},\tilde{T}\}$, is given by
\begin{equation}
A=\omega \tilde{J}+\omega ^{a}\tilde{G}_{a}+\tau \tilde{H}+e^{a}\tilde{P}%
_{a}+k\tilde{Z}+k^{a}\tilde{Z}_{a}+m\tilde{M}+s\tilde{S}+t\tilde{T}\,. \label{NR1f}
\end{equation}%
The EEB curvature two-form is then
\begin{eqnarray}
F &=&R\left( \omega \right) \tilde{J}+R^{a}\left( \omega ^{b}\right) \tilde{G%
}_{a}+R\left( \tau \right) \tilde{H}+R^{a}\left( e^{b}\right) \tilde{P}%
_{a}+R\left( k\right) \tilde{Z}+R^{a}\left( k^{b}\right) \tilde{Z}%
_{a}+R\left( m\right) \tilde{M}  \notag \\
&&+R\left( s\right) \tilde{S}+R\left( t\right) \tilde{T}\,,
\end{eqnarray}%
with\footnote{In the sequel, we will omit the wedge product $\wedge$ between differential forms.}
\begin{eqnarray}
R\left( \omega \right) &=&d\omega \,,  \notag \\
R^{a}\left( \omega ^{b}\right) &=&d\omega ^{a}+\epsilon ^{ac}\omega \omega
_{c}\,,  \notag \\
R\left( \tau \right) &=&d\tau \,,  \notag \\
R^{a}\left( e^{b}\right) &=&de^{a}+\epsilon ^{ac}\omega e_{c}+\epsilon
^{ac}\tau \omega _{c}+\frac{1}{\ell ^{2}}\epsilon ^{ac}ke_{c}+\frac{1}{\ell
^{2}}\epsilon ^{ac}\tau k_{c}\,,  \notag \\
R\left( k\right) &=&dk\,,  \notag \\
R^{a}\left( k^{b}\right) &=&dk^{a}+\epsilon ^{ac}\omega k_{c}+\epsilon
^{ac}\tau e_{c}+\epsilon ^{ac}k\omega _{c}+\frac{1}{\ell ^{2}}\epsilon
^{ac}kk_{c}\,,  \notag \\
R\left( m\right) &=&dm+\epsilon ^{ac}e_{a}\omega _{c}+\frac{1}{\ell ^{2}}%
\epsilon ^{ac}e_{a}k_{c}\,,  \notag \\
R\left( s\right) &=&ds+\frac{1}{2}\epsilon ^{ac}\omega _{a}\omega _{c}\,,
\notag \\
R\left( t\right) &=&dt+\epsilon ^{ac}\omega _{a}k_{c}+\frac{1}{2}\epsilon
^{ac}e_{a}e_{c}+\frac{1}{2\ell ^{2}}\epsilon ^{ac}k_{a}k_{c}\,.  \label{curvEEB}
\end{eqnarray}%
Then, the NR three-dimensional CS action based on the EEB algebra \cite{Concha:2019lhn}, up to boundary terms, is obtained considering the gauge connection one-form \eqref{NR1f} and the non-vanishing components of the invariant tensor \eqref{invt1} in the general expression of the CS action \eqref{genCS}. The NR CS graviy action reads
\begin{eqnarray}
I_{\text{EEB}} &=&\int \tilde{\alpha}_{0}\left[ \omega _{a}R^{a}\left(\omega ^{b}\right)-2sR ( \omega ) \right] +\tilde{\alpha}_{1}\left[
e_{a}R^{a}\left(\omega ^{b}\right)+ \omega_{a}R^{a}\left(e^{b}\right) -2mR(\omega )-2\tau ds \right.  \notag \\
&&+\left. \frac{1}{\ell ^{2}}e_{a}{R}^{a}\left(k^{b}\right) + \frac{1}{\ell ^{2}}k_{a} {R}^{a}\left(e^{b}\right) -\frac{2}{\ell ^{2}}\tau
dt -\frac{2}{\ell ^{2}}mR(k)\right] +\tilde{\alpha}_{2}\left[
e_{a}R^{a}\left( e^{b}\right) +k_{a}R^{a}\left( \omega ^{b}\right) \right.  \notag \\
&&+\left. \omega
_{a}R^{a}\left( k^{b}\right) + \frac{1}{\ell ^{2}}k_{a}R^{a}\left(k^{b}\right)-2sR( k) -2mR(
\tau) -2tR( \omega) -\frac{2}{\ell ^{2}}tR(k)\right] \,. \label{CS1}
\end{eqnarray}
The CS gravity theory based on the EEB symmetry can be seen as an alternative NR gravity model in presence of a cosmological constant. The NR CS action \eqref{CS1} contains three sectors proportional to different arbitrary
constants $\tilde{\alpha}_{0}$, $\tilde{\alpha}_{1}$, and $%
\tilde{\alpha}_{2}$. The first contribution, proportional to $\tilde{\alpha}%
_{0}$, is the CS action for the NR Exotic Gravity. The second and third
term, proportional, respectively, to $\tilde{\alpha}_{1}$ and $\tilde{\alpha}%
_{2}$, reproduce the enlarged extended Bargmann gravity with the explicit
presence of the $k_{a}$ gauge field. Observe that the limit $\ell
\rightarrow \infty $ taken in the term proportional to $\tilde{\alpha}_{1}$
reproduces the CS action for the Extended Bargmann algebra \cite{Bergshoeff:2016lwr}.
On the other hand, the limit $\ell \rightarrow \infty $ in the sector
proportional to $\tilde{\alpha}_{2}$ leads to the CS action for the NR
Maxwell algebra \cite{Aviles:2018jzw}. Let us note that the term proportional to $%
\tilde{\alpha}_{1}$ is not the extended Newton-Hooke gravity Lagrangian,
although it leads to the extended Bargmann gravity Lagrangian in the $\ell
\rightarrow \infty $ limit. In particular, the additional gauge fields
related to the EEB algebra appearing in the $\tilde{\alpha}_{1}$ term vanish
in the flat ($\ell \rightarrow \infty $) limit. 

\section{Three-dimensional supergravity based on a supersymmetric extension
of the Enlarged Extended Bargmann algebra}

Here, we present a supersymmetric extension of the EEB algebra which allows
us to construct a NR supergravity action. Consequently, we develop the
aforementioned NR supergravity action by exploiting the CS construction in
three dimensions. In order to have a proper NR CS supergravity action based
on a supersymmetric extension of the EEB algebra, one requires to find a NR
superalgebra which not only contains the EEB algebra as a subalgebra but
also admits an invariant supertrace. Such task can be properly accomplished by considering the $S$-expansion procedure \cite{Izaurieta:2006zz} (see also \cite{Ipinza:2016bfc, Penafiel:2016ufo} for recent developments on the same method; more details on the $S$-expansion procedure are given in Appendix \ref{appa}). Indeed, an EEB superalgebra can be obtained by applying the $S$-expansion to the $\mathcal{N}=2$ AdS-$\mathcal{L}$ superalgebra \cite{Concha:2019icz} with a suitable semigroup $S$.

\subsection{Supersymmetric extension of the Enlarged Extended Bargmann
algebra}
The relativistic $\mathcal{N}=2$ AdS-$\mathcal{L}$ superalgebra is characterized by the presence of the AdS-$\mathcal{L}$ bosonic generators $\{J_A,P_A,Z_A\}$, the fermionic spinor charges $\{Q_{\alpha}^{i},\Sigma_{\alpha}^{i}\}$, and three $\mathfrak{so}(2)$ generators $\mathcal{T}$, $\mathcal{U}$, and $\mathcal{B}$. Such generators satisfy the following non-vanishing commutation relations \cite{Concha:2019icz}:
\begin{eqnarray}
\left[ J_{A},J_{B}\right] &=&\epsilon _{ABC}J^{C}\,,\qquad\qquad\,\left[ J_{A},P_{B}\right] =\epsilon _{ABC}P^{C}\,, \qquad \qquad \, \left[ J_{A},Z_{B}\right] =\epsilon _{ABC}Z^{C}\,,\notag \\
\left[ P_{A},P_{B}\right] &=&\epsilon _{ABC}Z^{C}\,, \qquad \qquad \left[ Z_{A},Z_{B}\right] =\frac{1}{\ell ^{2}}\epsilon _{ABC}Z^{C}\,, \qquad \quad \left[ P_{A},Z_{B}\right] =\frac{1}{\ell ^{2}}\epsilon _{ABC}P^{C}\,, \notag \\
\left[J_{A},Q_{\alpha}^{i}\right] &=& -\frac{1}{2} \left( \gamma_{A} \right)_{\alpha}^{\ \beta} Q_{\beta}^{i}\,, \quad \ \ \left[J_{A},\Sigma_{\alpha}^{i}\right] = -\frac{1}{2} \left( \gamma_{A} \right)_{\alpha}^{\ \beta} \Sigma_{\beta}^{i}\,, \quad \ \ \left[P_{A},Q_{\alpha}^{i}\right] = -\frac{1}{2} \left( \gamma_{A} \right)_{\alpha}^{\ \beta} \Sigma_{\beta}^{i}\,,  \notag \\
\left[P_{A},\Sigma_{\alpha}^{i}\right] &=& -\frac{1}{2\ell^{2}} \left( \gamma_{A} \right)_{\alpha}^{\ \beta} Q_{\beta}^{i}\,, \ \ \left[Z_{A},Q_{\alpha}^{i}\right] = -\frac{1}{2\ell^{2}} \left( \gamma_{A} \right)_{\alpha}^{\ \beta} Q_{\beta}^{i}\,,  \ \ \, \left[Z_{A},\Sigma_{\alpha}^{i}\right] = -\frac{1}{2\ell^{2}} \left( \gamma_{A} \right)_{\alpha}^{\ \beta} \Sigma_{\beta}^{i}\,, \notag \\
 \left[\mathcal{T},Q_{\alpha}^{i}\right] &=& \frac{1}{2} \epsilon^{ij} Q_{\alpha}^{j}\,, \qquad \qquad \ \ \,  \left[\mathcal{T},\Sigma_{\alpha}^{i}\right] = \frac{1}{2} \epsilon^{ij} \Sigma_{\alpha}^{j}\,, \qquad \qquad  \ \ \ 
\left[\mathcal{U},Q_{\alpha}^{i}\right] = \frac{1}{2} \epsilon^{ij} \Sigma_{\alpha}^{j}\,, \notag \\
\left[\mathcal{U},\Sigma_{\alpha}^{i}\right] &=& \frac{1}{2\ell^{2}} \epsilon^{ij} Q_{\alpha}^{j}\,, \qquad \qquad  \left[\mathcal{B},Q_{\alpha}^{i}\right] = \frac{1}{2\ell^{2}} \epsilon^{ij} Q_{\alpha}^{j}\,, \qquad \qquad \, \left[\mathcal{B},\Sigma_{\alpha}^{i}\right] = \frac{1}{2\ell^{2}} \epsilon^{ij} \Sigma_{\alpha}^{j}\,, \label{n2SADSL}
\end{eqnarray}
along with the following anti-commutators,
\begin{eqnarray}
\{ Q_{\alpha}^{i},Q_{\beta}^{j}\} &=&-\delta_{ij}\left(\gamma^{A}C \right)_{\alpha \beta} P_{A}-C_{\alpha \beta}\epsilon^{ij}\mathcal{U}\,, \notag\\ 
\{ Q_{\alpha}^{i},\Sigma_{\beta}^{j}\} &=&-\delta_{ij}\left(\gamma^{A}C \right)_{\alpha \beta} Z_{A}-C_{\alpha \beta}\epsilon^{ij}\mathcal{B}\,, \notag\\
\{ \Sigma_{\alpha}^{i},\Sigma_{\beta}^{j}\} &=&-\frac{\delta_{ij}}{\ell^{2}}\left(\gamma^{A}C \right)_{\alpha \beta} P_{A}-\frac{1}{\ell^{2}}C_{\alpha \beta}\epsilon^{ij}\mathcal{U}\,, \label{n2SADSLa}
\end{eqnarray}
where $A,B,C=0,1,2$ are the Lorentz indices which are raised and lowered with the Minkoswki metric $\eta_{AB}=(-1,1,1)$, $\alpha,\beta=1,2$,\footnote{In the following, we will frequently omit these spinorial indices in order to lighten the notation.} and $i,j=1,2$ label the number of supercharges. Here, $\gamma^{A}$ are Dirac matrices in three spacetime dimensions and $C$ is the charge conjugation matrix,
\begin{equation}
C_{\alpha\beta}=C^{\alpha\beta}=\begin{pmatrix}
0 & -1\\
1 & 0\\
\end{pmatrix}\, ,
\end{equation}
which satisfies $C^{T}=-C$ and $C\gamma^{A}=(C\gamma^{A})^{T}$. As it was discussed in \cite{Concha:2019icz}, the presence of $\mathfrak{so}(2)$ symmetry generators is not arbitrary and assures the non-degeneracy of the invariant tensor. In particular, the $\mathcal{N}=2$ AdS-$\mathcal{L}$ superalgebra admits the following non-vanishing components of an invariant tensor,
\begin{eqnarray}
\langle J_A J_B \rangle &=&\alpha_0 \eta_{AB}\,, \qquad \qquad \langle J_A P_B \rangle =\alpha_1 \eta_{AB}\,, \qquad \qquad \langle J_A Z_B \rangle =\alpha_2 \eta_{AB}\,, \notag \\
\langle P_A P_B \rangle &=&\alpha_2 \eta_{AB}\,, \qquad \quad \ \ \, \langle P_A Z_B \rangle =\frac{\alpha_1}{\ell^{2}} \eta_{AB}\,, \qquad \quad \ \ \langle Z_A Z_B \rangle =\frac{\alpha_2}{\ell^{2}} \eta_{AB}\,, \notag \\
\langle \mathcal{T T} \rangle&=&\alpha_0\,, \qquad \qquad \qquad \ \ \langle \mathcal{T U} \rangle=\alpha_1\,, \qquad \qquad \qquad \ \ \,  \langle \mathcal{T B} \rangle=\alpha_2\,, \qquad \qquad \notag \\
\langle \mathcal{U U} \rangle&=&\alpha_2\,, \qquad \qquad \qquad \ \ \,\langle \mathcal{U B} \rangle=\frac{\alpha_1}{\ell^{2}}\,, \qquad \qquad \qquad \ \  \langle \mathcal{B B} \rangle=\frac{\alpha_2}{\ell^{2}}\,, \notag \\
\langle Q_{\alpha}^{i} Q_{\beta}^{j} \rangle&=&\alpha_1 C_{\alpha\beta}\delta^{ij}\,,\qquad \ \ \, \langle Q_{\alpha}^{i} \Sigma_{\beta}^{j} \rangle=\alpha_2 C_{\alpha\beta}\delta^{ij}\,,\qquad \ \ \, \langle \Sigma_{\alpha}^{i} \Sigma_{\beta}^{j} \rangle=\frac{\alpha_1}{\ell^{2}} C_{\alpha\beta}\delta^{ij}\,. \label{relinvt}
\end{eqnarray}
One can notice that the flat limit $\ell\rightarrow\infty$ leads to the $\mathcal{N}=2$ Maxwell supergravity theory \cite{Concha:2019icz} in which the internal symmetry generator $\mathcal{B}$ becomes a central charge.

A supersymmetric extension of the EEB algebra can be obtained by expanding the $\mathcal{N}=2$ AdS-$\mathcal{L}$ superalgebra \eqref{n2SADSL}-\eqref{n2SADSLa}. Indeed, as we shall see, the $S$-expansion of a relativistic superalgebra considering $S_{E}^{(2)}$ as the relevant semigroup reproduces a NR contraction. This is due to the fact that the $S$-expansion method can be seen as a generalization of the Inönü-Wigner contraction process \cite{Inonu:1953sp}, when the semigroup under consideration belongs to the $S_{E}^{(N)}$ family. In particular, it would be interesting to analyze which kind of superalgebras can be found by considering $S_{E}
^{(N)}$ for $N>2$. Here, we shall focus on the semigroup $S_{E}^{(2)}$, which allows us to obtain not only a well-defined EEB superalgebra but also its non-degenerate invariant tensor.

Let $S_{E}^{(2)}=\{ \lambda_0,\lambda_1,\lambda_2,\lambda_3 \}$ be the relevant semigroup whose elements satisfy the following multiplication law:
\begin{equation}
\begin{tabular}{l|llll}
$\lambda _{3}$ & $\lambda _{3}$ & $\lambda _{3}$ & $\lambda _{3}$ & $\lambda
_{3}$ \\
$\lambda _{2}$ & $\lambda _{2}$ & $\lambda _{3}$ & $\lambda _{3}$ & $\lambda
_{3}$ \\
$\lambda _{1}$ & $\lambda _{1}$ & $\lambda _{2}$ & $\lambda _{3}$ & $\lambda
_{3}$ \\
$\lambda _{0}$ & $\lambda _{0}$ & $\lambda _{1}$ & $\lambda _{2}$ & $\lambda
_{3}$ \\ \hline
& $\lambda _{0}$ & $\lambda _{1}$ & $\lambda _{2}$ & $\lambda _{3}$%
\end{tabular}
\label{ml}
\end{equation}%
Here $\lambda_3=0_s$ is the zero element of the semigroup such that $0_s\lambda_k=0_s$. Before applying the $S$-expansion to the $\mathcal{N}=2$ AdS-$\mathcal{L}$ superalgebra we will consider a particular subspace decomposition of the Lie superalgebra. Let $V_0=\{J_0,P_0,Z_0,\mathcal{T},\mathcal{U},\mathcal{B},Q_{\alpha}^{+},\Sigma_{\alpha}^{+} \}$ and $V_1=\{J_a,P_a,Z_a,Q_{\alpha}^{-},\Sigma_{\alpha}
^{-}\}$ be the subspaces decomposition of the $\mathcal{N}=2$ superalgebra \eqref{n2SADSL}-\eqref{n2SADSLa}, where we have split the Lorentz index as $A=\lbrace 0,a \rbrace$ with $a=1,2$ and where we have defined
\begin{equation}
    Q_{\alpha}^{\pm}=\frac{1}{\sqrt{2}}\left( Q_{\alpha}^{1}\pm \epsilon_{\alpha\beta}Q_{\beta}^{2}\right)\,, \qquad 
    \Sigma_{\alpha}^{\pm}=\frac{1}{\sqrt{2}}\left( \Sigma_{\alpha}^{1}\pm \epsilon_{\alpha\beta}\Sigma_{\beta}^{2}\right)\,.\label{redeff}
\end{equation}
One can see that such decomposition satisfies
\begin{eqnarray}
[V_0,V_0]\subset V_0\,,\qquad [V_0,V_1]\subset V_1\,, \qquad [V_1,V_1]\subset V_0\,.
\end{eqnarray}
Let us consider now $S_{E}^{(2)}=S_{0} \cup S_{1}$ as the semigroup decomposition where
\begin{eqnarray}
S_0&=&\{\lambda_0,\lambda_2,\lambda_3\}\,, \notag \\
S_1&=&\{\lambda_1,\lambda_3\} \,. 
\label{decomp}
\end{eqnarray}
Then, the decomposition \eqref{decomp} is said to be resonant since it satisfies the same structure than the subspaces, that is
\begin{equation}
    S_0\cdot S_0\subset S_0\,,\qquad \ S_0\cdot S_1\subset S_1\,,\qquad \ S_1\cdot S_1\subset S_0\,.\label{semidecomp}
\end{equation}
Following the definitions of \cite{Izaurieta:2006zz}, after extracting a resonant subalgebra of the $S_{E}^{(2)}$-expansion of the $\mathcal{N}=2$ AdS-$\mathcal{L}$ superalgebra and applying a $0_s$-reduction, one finds a novel NR expanded superalgebra spanned by the set of generators: 
\begin{equation}
\{\tilde{J},\tilde{G}_{a},\tilde{S},\tilde{H},\tilde{P}_{a},\tilde{M},\tilde{%
Z},\tilde{Z}_{a},\tilde{T},\tilde{Y}_{1},\tilde{Y}_{2},\tilde{U}_{1},\tilde{U%
}_{2},\tilde{B}_{1},\tilde{B}_{2},\tilde{Q}_{\alpha }^{+},\tilde{Q}_{\alpha
}^{-},\tilde{R}_{\alpha },\tilde{\Sigma}_{\alpha }^{+},\tilde{\Sigma}%
_{\alpha }^{-},\tilde{W}_{\alpha }\}\,.
\end{equation}%
The NR generators are related to the relativistic ones through the semigroup elements as
\begin{eqnarray}
\tilde{J}&=&\lambda_0 J_0\,, \qquad  \ \, \tilde{S}=\lambda_2 J_0\,, \qquad \tilde{G}_a=\lambda_1 J_a\,, \notag \\
\tilde{H}&=&\lambda_0 P_0\,, \qquad \tilde{M}=\lambda_2 P_0\,, \qquad \tilde{P}_a=\lambda_1 P_a \,, \notag \\
\tilde{Z}&=&\lambda_0 Z_0\,, \qquad \ \tilde{T}=\lambda_2 Z_0\,, \qquad \tilde{Z}_a=\lambda_1 Z_a\,, \notag \\
\tilde{Q}_{\alpha}^{+}&=&\lambda_0 Q_{\alpha}^{+}\,, \quad \ \, \tilde{R}_{\alpha}=\lambda_2 Q_{\alpha}^{+}\,, \quad \ \tilde{Q}_{\alpha}^{-}=\lambda_1 Q_{\alpha}^{-}\,, \notag \\
\tilde{\Sigma}_{\alpha}^{+}&=&\lambda_0 \Sigma_{\alpha}^{+}\,, \quad \  \tilde{W}_{\alpha}=\lambda_2 \Sigma_{\alpha}^{+}\,, \quad \ \, \tilde{\Sigma}_{\alpha}^{-}=\lambda_1 \Sigma_{\alpha}^{-}\,, \notag \\
\tilde{Y}_{1}&=&\lambda_0 \mathcal{T}\,, \qquad \  \tilde{U}_1=\lambda_0 \mathcal{U}\,, \qquad \  \tilde{B}_1=\lambda_0 \mathcal{B}\,, \notag\\
\tilde{Y}_{2}&=&\lambda_2 \mathcal{T}\,, \qquad \  \tilde{U}_2=\lambda_2 \mathcal{U}\,, \qquad \  \tilde{B}_2=\lambda_2 \mathcal{B}\,.\label{sexp1}
\end{eqnarray}
The NR generators satisfy the bosonic subalgebra \eqref{EEB1} along with the following commutation relations:
\begin{eqnarray}
\left[ \tilde{J},\tilde{Q}_{\alpha }^{\pm }\right] &=&-\frac{1}{2}\left(
\gamma _{0}\right) _{\alpha }^{\text{ }\beta }\tilde{Q}_{\beta }^{\pm
}\,,\qquad \left[ \tilde{J},\tilde{R}_{\alpha }\right] =-%
\frac{1}{2}\left( \gamma _{0}\right) _{\alpha }^{\text{ }\beta }\tilde{R%
}_{\beta }\,,   \qquad \, \left[ \tilde{H},\tilde{Q}_{\alpha }^{\pm}\right] =-\frac{1}{2}\left(
\gamma _{0}\right) _{\alpha }^{\text{ }\beta }\tilde{\Sigma}_{\beta }^{\pm
}\,,  \notag\\
\left[ \tilde{J},\tilde{\Sigma}_{\alpha }^{\pm }\right] &=&-\frac{1}{2}\left(
\gamma _{0}\right) _{\alpha }^{\text{ }\beta }\tilde{\Sigma}_{\beta }^{\pm
}\,,\qquad \left[ \tilde{J},\tilde{W}_{\alpha }\right] =-%
\frac{1}{2}\left( \gamma _{0}\right) _{\alpha }^{\text{ }\beta }\tilde{W%
}_{\beta }\,,   \quad \ \ \, \left[ \tilde{H},\tilde{\Sigma}_{\alpha }^{\pm}\right] =-\frac{1}{2\ell^{2}}\left(
\gamma _{0}\right) _{\alpha }^{\text{ }\beta }\tilde{Q}_{\beta}^{\pm}\,,  \notag\\
\left[ \tilde{Z},\tilde{Q}_{\alpha }^{\pm}\right] &=&-\frac{1}{2\ell^{2}}\left(
\gamma _{0}\right) _{\alpha }^{\text{ }\beta }\tilde{Q}_{\beta }^{\pm
}\,,  \quad  \left[ \tilde{H},\tilde{R}_{\alpha }\right] =-\frac{1}{2}\left(
\gamma _{0}\right) _{\alpha }^{\text{ }\beta }\tilde{W}_{\beta }\,, \quad \ \ \ \left[ \tilde{Z},\tilde{\Sigma}_{\alpha }^{\pm}\right] =-\frac{1}{2\ell^{2}}\left(
\gamma _{0}\right) _{\alpha }^{\text{ }\beta }\tilde{\Sigma}_{\beta }^{\pm}\,, \notag \\
\left[ \tilde{S},\tilde{Q}_{\alpha }^{+}\right] &=&-\frac{1}{2}\left(
\gamma _{0}\right) _{\alpha }^{\text{ }\beta }\tilde{R}_{\beta }\,, \quad \ \
\left[ \tilde{M},\tilde{Q}_{\alpha }^{+}\right] =-\frac{1}{2}\left(
\gamma _{0}\right) _{\alpha }^{\text{ }\beta }\tilde{W}_{\beta }\,, \quad \ \ \ \, \left[ \tilde{T},\tilde{Q}_{\alpha }^{+}\right] =-\frac{1}{2\ell^{2}}\left(
\gamma _{0}\right) _{\alpha }^{\text{ }\beta }\tilde{R}_{\beta }\,, \notag \\
\left[ \tilde{S},\tilde{\Sigma}_{\alpha }^{+}\right] &=&-\frac{1}{2}\left(
\gamma _{0}\right) _{\alpha }^{\text{ }\beta }\tilde{W}_{\beta }\,, \quad \ \,
\left[ \tilde{M},\tilde{\Sigma}_{\alpha }^{+}\right] =-\frac{1}{2\ell^{2}}\left(
\gamma _{0}\right) _{\alpha }^{\text{ }\beta }\tilde{R}_{\beta }\,, \quad \ \,  \left[ \tilde{T},\tilde{\Sigma}_{\alpha }^{+}\right] =-\frac{1}{2\ell^{2}}\left(
\gamma _{0}\right) _{\alpha }^{\text{ }\beta }\tilde{W}_{\beta }\,, \notag \\
\left[ \tilde{H},\tilde{W}_{\alpha }\right] &=&-\frac{1}{2\ell^{2}}\left(
\gamma _{0}\right) _{\alpha }^{\text{ }\beta }\tilde{R}_{\beta }\,, \quad \,
\left[ \tilde{Z},\tilde{R}_{\alpha }\right] =-\frac{1}{2\ell^{2}}\left(
\gamma _{0}\right) _{\alpha }^{\text{ }\beta }\tilde{R}_{\beta }\,, \quad \   \left[ \tilde{Z},\tilde{W}_{\alpha }\right] =-\frac{1}{2\ell^{2}}\left(
\gamma _{0}\right) _{\alpha }^{\text{ }\beta }\tilde{W}_{\beta }\,, \notag \\
\left[ \tilde{G}_{a},\tilde{Q}_{\alpha }^{+}\right] &=&-\frac{1}{2}\left(
\gamma _{a}\right) _{\alpha }^{\text{ }\beta }\tilde{Q}_{\beta
}^{-}\,, \quad \ 
\left[ \tilde{G}_{a},\tilde{Q}_{\alpha }^{-}\right] =-\frac{%
1}{2}\left( \gamma _{a}\right) _{\alpha }^{\text{ }\beta }\tilde{R}_{\beta
}\,, \quad \ \ \, \left[ \tilde{P}_{a},\tilde{Q}_{\alpha }^{+}\right] =-\frac{%
1}{2}\left( \gamma _{a}\right) _{\alpha }^{\text{ }\beta }\tilde{\Sigma}_{\beta
}^{-}\,, \notag \\
\left[ \tilde{G}_{a},\tilde{\Sigma}_{\alpha }^{+}\right] &=&-\frac{1}{2}\left(
\gamma _{a}\right) _{\alpha }^{\text{ }\beta }\tilde{\Sigma}_{\beta
}^{-}\,, \quad \ 
\left[ \tilde{G}_{a},\tilde{\Sigma}_{\alpha }^{-}\right] =-\frac{%
1}{2}\left( \gamma _{a}\right) _{\alpha }^{\text{ }\beta }\tilde{W}_{\beta
}\,, \quad \ \ \, \left[ \tilde{P}_{a},\tilde{Q}_{\alpha }^{-}\right] =-\frac{%
1}{2}\left( \gamma _{a}\right) _{\alpha }^{\text{ }\beta }\tilde{W}_{\beta
}^{-}\,, \notag \\
\left[ \tilde{P}_{a},\tilde{\Sigma}_{\alpha }^{+}\right] &=&-\frac{%
1}{2\ell^{2}}\left( \gamma _{a}\right) _{\alpha }^{\text{ }\beta }\tilde{Q}_{\beta
}^{-}\,, \  \, 
\left[ \tilde{Z}_{a},\tilde{Q}_{\alpha }^{+}\right] =-\frac{%
1}{2\ell^{2}}\left( \gamma _{a}\right) _{\alpha }^{\text{ }\beta }\tilde{Q}_{\beta
}^{-}\,, \ \ \ \left[ \tilde{Z}_{a},\tilde{Q}_{\alpha }^{-}\right] =-\frac{%
1}{2\ell^{2}}\left( \gamma _{a}\right) _{\alpha }^{\text{ }\beta }\tilde{R}_{\beta
}\,,  \notag \\
\left[ \tilde{P}_{a},\tilde{\Sigma}_{\alpha }^{-}\right] &=&-\frac{%
1}{2\ell^{2}}\left( \gamma _{a}\right) _{\alpha }^{\text{ }\beta }\tilde{R}_{\beta
}\,, \ \ \, 
\left[ \tilde{Z}_{a},\tilde{\Sigma}_{\alpha }^{+}\right] =-\frac{%
1}{2\ell^{2}}\left( \gamma _{a}\right) _{\alpha }^{\text{ }\beta }\tilde{\Sigma}_{\beta
}^{-}\,, \ \ \ \left[ \tilde{Z}_{a},\tilde{\Sigma}_{\alpha }^{-}\right] =-\frac{%
1}{2\ell^{2}}\left( \gamma _{a}\right) _{\alpha }^{\text{ }\beta }\tilde{W}_{\beta
}\,, \notag \\ \label{SEEBa}
\end{eqnarray}
while the additional bosonic generators $\{Y_1,Y_2,U_1,U_2,B_1,B_2\}$, which are expansions of $\mathfrak{so}(2)$ internal symmetry generators, satisfy the following commutators:
\begin{eqnarray}
\left[ \tilde{Y}_{1},\tilde{Q}_{\alpha }^{+}\right] &=&\frac{1}{2}\left(
\gamma _{0}\right) _{\alpha \beta }\tilde{Q}_{\beta }^{+}\,,\qquad \left[ \tilde{Y}_{1},\tilde{Q}_{\alpha }^{-}\right] =-\frac{1}{2}\left(
\gamma _{0}\right) _{\alpha \beta }\tilde{Q}_{\beta }^{-}\,, \quad \ \ \,   \left[ \tilde{Y}_{1},\tilde{R}_{\alpha }\right] =\frac{1}{2}\left(
\gamma _{0}\right) _{\alpha \beta }\tilde{R}_{\beta }\,,  \notag\\
\left[ \tilde{Y}_{1},\tilde{\Sigma}_{\alpha }^{+}\right] &=&\frac{1}{2}\left(
\gamma _{0}\right) _{\alpha \beta }\tilde{\Sigma}_{\beta }^{+}\,,\qquad \, \left[ \tilde{Y}_{1},\tilde{\Sigma}_{\alpha }^{-}\right] =-\frac{1}{2}\left(
\gamma _{0}\right) _{\alpha \beta }\tilde{\Sigma}_{\beta }^{-}\,, \quad \ \   \left[ \tilde{Y}_{1},\tilde{W}_{\alpha }\right] =\frac{1}{2}\left(
\gamma _{0}\right) _{\alpha \beta }\tilde{W}_{\beta }\,,  \notag\\
\left[ \tilde{Y}_{2},\tilde{Q}_{\alpha }^{+}\right] &=&\frac{1}{2}\left(
\gamma _{0}\right) _{\alpha \beta }\tilde{R}_{\beta }\,, \qquad \, \left[ \tilde{U}_{1},\tilde{Q}_{\alpha }^{-}\right] =-\frac{1}{2}\left(
\gamma _{0}\right) _{\alpha \beta }\tilde{\Sigma}_{\beta }^{-}\,,  \quad \ \, \left[ \tilde{U}_{1},\tilde{Q}_{\alpha }^{+}\right] =\frac{1}{2}\left(
\gamma _{0}\right) _{\alpha \beta }\tilde{\Sigma}_{\beta }^{+}\,,  \notag\\
\left[ \tilde{Y}_{2},\tilde{\Sigma}_{\alpha }^{+}\right] &=&\frac{1}{2}\left(
\gamma _{0}\right) _{\alpha \beta }\tilde{W}_{\beta }\, \qquad \  \left[ \tilde{U}_{1},\tilde{R}_{\alpha }\right] =\frac{1}{2}\left(
\gamma _{0}\right) _{\alpha \beta }\tilde{W}_{\beta }\,, \qquad \  \left[ \tilde{U}_{2},\tilde{Q}_{\alpha }^{+}\right] =\frac{1}{2}\left(
\gamma _{0}\right) _{\alpha \beta }\tilde{W}_{\beta }\,,  \notag \\
\left[ \tilde{U}_{1},\tilde{\Sigma}_{\alpha }^{+}\right] &=&\frac{1}{2\ell^{2}}\left(
\gamma _{0}\right) _{\alpha \beta }\tilde{Q}_{\beta }^{+}\,,\quad \, \left[ \tilde{U}_{1},\tilde{\Sigma}_{\alpha }^{-}\right] =-\frac{1}{2\ell^{2}}\left(
\gamma _{0}\right) _{\alpha \beta }\tilde{Q}_{\beta }^{-}\,, \ \    \left[ \tilde{B}_{1},\tilde{R}_{\alpha }\right] =\frac{1}{2\ell^{2}}\left(
\gamma _{0}\right) _{\alpha \beta }\tilde{R}_{\beta }\,,  \notag\\
\left[ \tilde{B}_{1},\tilde{Q}_{\alpha }^{+}\right] &=&\frac{1}{2\ell^{2}}\left(
\gamma _{0}\right) _{\alpha \beta }\tilde{Q}_{\beta }^{+}\,,\quad  \left[ \tilde{B}_{1},\tilde{Q}_{\alpha }^{-}\right] =-\frac{1}{2\ell^{2}}\left(
\gamma _{0}\right) _{\alpha \beta }\tilde{Q}_{\beta }^{-}\,, \ \,    \left[ \tilde{B}_{1},\tilde{W}_{\alpha }\right] =\frac{1}{2\ell^{2}}\left(
\gamma _{0}\right) _{\alpha \beta }\tilde{W}_{\beta }\,,  \notag\\
\left[ \tilde{B}_{1},\tilde{\Sigma}_{\alpha }^{+}\right] &=&\frac{1}{2\ell^{2}}\left(
\gamma _{0}\right) _{\alpha \beta }\tilde{\Sigma}_{\beta }^{+}\,,\quad \, \left[ \tilde{B}_{1},\tilde{\Sigma}_{\alpha }^{-}\right] =-\frac{1}{2\ell^{2}}\left(
\gamma _{0}\right) _{\alpha \beta }\tilde{\Sigma}_{\beta }^{-}\,, \ \    \left[ \tilde{U}_{1},\tilde{W}_{\alpha }\right] =\frac{1}{2\ell^{2}}\left(
\gamma _{0}\right) _{\alpha \beta }\tilde{R}_{\beta }\,,  \notag\\
\left[ \tilde{U}_{2},\tilde{\Sigma}_{\alpha }^{+}\right] &=&\frac{1}{2\ell^{2}}\left(
\gamma _{0}\right) _{\alpha \beta }\tilde{R}_{\beta }\,,\quad \, \left[ \tilde{B}_{2},\tilde{Q}_{\alpha }^{+}\right] =\frac{1}{2\ell^{2}}\left(
\gamma _{0}\right) _{\alpha \beta }\tilde{R}_{\beta }\,, \quad \ \    \left[ \tilde{B}_{2},\tilde{\Sigma}_{\alpha }^{+}\right] =\frac{1}{2\ell^{2}}\left(
\gamma _{0}\right) _{\alpha \beta }\tilde{W}_{\beta }\,. \notag \\
\label{SEEBb}
\end{eqnarray}
On the other hand, the fermionic generators satisfy the following anti-commutation relations:
\begin{eqnarray}
\left\{ \tilde{Q}_{\alpha }^{-},\tilde{Q}_{\beta }^{-}\right\} &=&-\left(
\gamma ^{0}C\right) _{\alpha \beta }\tilde{M}+\left( \gamma
^{0}C\right) _{\alpha \beta }\tilde{U}_{2}\,, \qquad
\left\{ \tilde{Q}_{\alpha }^{+},\tilde{Q}_{\beta }^{+}\right\} =-\left(
\gamma ^{0}C\right) _{\alpha \beta }\tilde{H}-\left( \gamma ^{0}C\right)
_{\alpha \beta }\tilde{U}_{1}\,,  \notag \\
\left\{ \tilde{Q}_{\alpha }^{-},\tilde{\Sigma}_{\beta }^{-}\right\}
&=&-\left( \gamma ^{0}C\right) _{\alpha \beta }\tilde{T}+%
\left( \gamma ^{0}C\right) _{\alpha \beta }\tilde{B}_{2}\,, \qquad \
\left\{ \tilde{Q}_{\alpha }^{+},\tilde{\Sigma}_{\beta }^{+}\right\}
=-\left( \gamma ^{0}C\right) _{\alpha \beta }\tilde{Z}-\left( \gamma
^{0}C\right) _{\alpha \beta }\tilde{B}_{1}\,,  \notag \\
\left\{ \tilde{Q}_{\alpha }^{+},\tilde{R}_{\beta }\right\} &=&-\left( \gamma
^{0}C\right) _{\alpha \beta }\tilde{M}-\left( \gamma ^{0}C\right) _{\alpha
\beta }\tilde{U}_{2}\,,   \qquad
\left\{ \tilde{Q}_{\alpha }^{+},\tilde{W}_{\beta }\right\} =-\left( \gamma
^{0}C\right) _{\alpha \beta }\tilde{T}-\left( \gamma ^{0}C\right) _{\alpha
\beta }\tilde{B}_{2}\,,  \notag \\
\left\{ \tilde{\Sigma}_{\alpha }^{+},\tilde{R}_{\beta }\right\} &=&-\left(
\gamma ^{0}C\right) _{\alpha \beta }\tilde{T}-\left( \gamma ^{0}C\right)
_{\alpha \beta }\tilde{B}_{2}\,, \qquad \ \,
\left\{ \tilde{\Sigma}_{\alpha }^{-},\tilde{\Sigma}_{\beta }^{-}\right\} =-\left( \gamma ^{0}C\right) _{\alpha \beta }\frac{\tilde{M}}{\ell ^{2}}%
+\left( \gamma ^{0}C\right) _{\alpha
\beta }\frac{\tilde{U}_{2}}{\ell ^{2}}\,,  \notag \\
\left\{ \tilde{\Sigma}_{\alpha }^{+},\tilde{\Sigma}_{\beta }^{+}\right\} &=&-%
\left( \gamma ^{0}C\right) _{\alpha \beta }\frac{\tilde{H}}{\ell ^{2}}-%
\left( \gamma ^{0}C\right) _{\alpha \beta }\frac{\tilde{U}_1}{\ell ^{2}}%
\,, \quad \ \ \,
\left\{ \tilde{\Sigma}_{\alpha }^{+},\tilde{W}_{\beta }\right\} =-\left( \gamma ^{0}C\right) _{\alpha \beta }\frac{\tilde{M}}{%
\ell ^{2}}-\left( \gamma ^{0}C\right) _{\alpha \beta }\frac{\tilde{U}_{2}}{\ell
^{2}}\,,  \notag\\
\left\{ \tilde{Q}_{\alpha }^{+},\tilde{Q}_{\beta }^{-}\right\} &=&-\left(
\gamma ^{a}C\right) _{\alpha \beta }\tilde{P}_{a}\,, \qquad \qquad \qquad \qquad \ \, \left\{ \tilde{Q}_{\alpha }^{\pm },\tilde{\Sigma}_{\beta }^{\mp }\right\}
=-\left( \gamma ^{a}C\right) _{\alpha \beta }\tilde{Z}_{a}\,,\quad
\notag \\  
\left\{ \tilde{\Sigma}_{\alpha }^{+},\tilde{\Sigma}_{\beta }^{-}\right\} &=&-%
\frac{1}{\ell ^{2}}\left( \gamma ^{a}C\right) _{\alpha \beta }\tilde{P}%
_{a}\,.  \label{SEEBc}
\end{eqnarray}%
The superalgebra given by \eqref{EEB1}, \eqref{SEEBa}, \eqref{SEEBb}, and \eqref{SEEBc} will be
denoted as the \textit{enlarged Extended Bargmann} superalgebra and, as we can see,
it properly contains the EEB algebra \cite{Concha:2018jjj} as bosonic subalgebra. 
 Let us note that the presence of the $\tilde{R}$ and $\tilde{W}$
generators is similar to what happens in the Extended Bargmann superalgebra
presented in \cite{Bergshoeff:2016lwr} and the Extended Newtonian superalgebra of \cite{Ozdemir:2019tby}
in which a $\tilde{R}$ generator is considered. One can see that the supersymmetric extension of the EEB algebra requires the presence of six additional bosonic generators $\tilde{Y}_{1}$, $\tilde{Y}_{2}$, $\tilde{U}_{1}$, $\tilde{U}_{2}$, $\tilde{B}_{1}$, and $\tilde{B}_{2}$ which act non-trivially on the fermionic charges $\tilde{Q}_{\alpha}^{\pm}$, $\tilde{\Sigma}_{\alpha}
^{\pm}$, $\tilde{R}_{\alpha}$, and $\tilde{W}_{\alpha}$. Interestingly, both $\tilde{B}_1$ and $\tilde{B}_2$ become central in the vanishing cosmological constant limit.
Indeed, the flat limit $\ell \rightarrow \infty $ of the EEB superalgebra reproduces the Maxwellian version of the extended Bargmann superalgebra introduced in \cite{Concha:2019mxx}, and corresponds to the supersymmetric extension of the MEB algebra first presented in \cite{Aviles:2018jzw}. Let us stress that the EEB superalgebra obtained
here has been obtained through an $S_{E}^{(2)}$-expansion of a relativistic
superalgebra. As we shall see in the next section, the supersymmetric extension of the EEB algebra presented here allowing the construction a well-defined CS supergravity action is not unique. Indeed, a different supersymmetric extension of the AdS-$\mathcal{L}$ algebra can be expanded with the same semigroup $S_{E}^{(2)}$ to get a different EEB superalgebra. 

Before studying the construction of a NR CS supergravity action invariant under the novel NR superalgebra, we will show that the present NR structure can be rewritten in a different way. Indeed, the Nappi-Witten symmetry \cite{Nappi:1993ie,Figueroa-OFarrill:1999cmq}, which can be seen as a central extension of the homogeneous part of the Galilei algebra, appears considering an appropriate redefinition of the generators. In particular, the EEB superalgebra can be written as three copies of the Nappi-Witten algebra, two of which are augmented by supersymmetry, endowed with $\mathfrak{u}(1)$ generators. Indeed, the aforesaid structure is revealed by considering the following redefinition:
\begin{eqnarray}
G_a&=&\frac{1}{2}\left( \ell^{2}\tilde{Z}_{a}+\ell\tilde{P}_{a} \right)\,, \qquad \quad G_{a}^{*}=\frac{1}{2}\left( \ell^{2}\tilde{Z}_{a}-\ell\tilde{P}_{a} \right)\,, \qquad \quad \hat{G}_{a}=\tilde{G}_{a}-\ell^{2}\tilde{Z}_{a}\,, \notag \\
J&=&\frac{1}{2}\left( \ell^{2}\tilde{Z}+\ell\tilde{H} \right)\,, \qquad \quad \ \ \, J^{*}=\frac{1}{2}\left( \ell^{2}\tilde{Z}-\ell\tilde{H} \right)\,, \qquad \quad \ \ \ \, \hat{J}=\tilde{J}-\ell^{2}\tilde{Z}\,, \notag \\
S&=&\frac{1}{2}\left( \ell^{2}\tilde{T}+\ell\tilde{M} \right)\,, \qquad \quad \ \ S^{*}=\frac{1}{2}\left( \ell^{2}\tilde{T}-\ell\tilde{M} \right)\,, \qquad \quad \ \ \ \hat{S}=\tilde{S}-\ell^{2}\tilde{T}\,, \notag \\
T_{1}&=&\frac{1}{2}\left( \ell^{2}\tilde{B}_{1}+\ell\tilde{U}_{1} \right)\,, \qquad \quad \, T_{1}^{*}=\frac{1}{2}\left( \ell^{2}\tilde{B}_{1}-\ell\tilde{U}_{1} \right)\,, \qquad \quad \, \hat{T}_{1}=\tilde{Y}_{1}-\ell^{2}\tilde{B}_{1}\,, \notag \\
T_{2}&=&\frac{1}{2}\left( \ell^{2}\tilde{B}_{2}+\ell\tilde{U}_{2} \right)\,, \qquad \quad \, T_{2}^{*}=\frac{1}{2}\left( \ell^{2}\tilde{B}_{2}-\ell\tilde{U}_{2} \right)\,, \qquad \quad \, \hat{T}_{2}=\tilde{Y}_{2}-\ell^{2}\tilde{B}_{2}\,, \notag \\
\mathcal{Q}_{\alpha}^{+}&=&\frac{1}{2}\left(\ell^{1/2}\tilde{Q}_{\alpha}^{+}+\ell^{3/2}\tilde{\Sigma}_{\alpha}^{+} \right)\,, \ \mathcal{Q}_{\alpha}^{-}=\frac{1}{2}\left(\ell^{1/2}\tilde{Q}_{\alpha}^{-}+\ell^{3/2}\tilde{\Sigma}_{\alpha}^{-} \right)\,, \ \mathcal{R}_{\alpha}=\frac{1}{2}\left(\ell^{1/2}\tilde{R}_{\alpha}+\ell^{3/2}\tilde{W}_{\alpha} \right)\,, \notag \\
\mathcal{G}_{\alpha}^{+}&=&\frac{i}{2}\left(\ell^{1/2}\tilde{Q}_{\alpha}^{+}-\ell^{3/2}\tilde{\Sigma}_{\alpha}^{+} \right)\,, \ \, \mathcal{G}_{\alpha}^{-}=\frac{i}{2}\left(\ell^{1/2}\tilde{Q}_{\alpha}^{-}-\ell^{3/2}\tilde{\Sigma}_{\alpha}^{-} \right)\,, \, \mathcal{W}_{\alpha}=\frac{i}{2}\left(\ell^{1/2}\tilde{R}_{\alpha}-\ell^{3/2}\tilde{W}_{\alpha} \right)\,. \notag \\
\label{redef1a}
\end{eqnarray}
One can notice that set of generators $\{ \hat{G}_{a},\hat{J},\hat{S},\hat{T}_{1},\hat{T}_{2} \}$ satisfies the Nappi-Witten algebra \cite{Nappi:1993ie} endowed with $\mathfrak{u}(1)$ generators $\hat{T}_{1}$ and $\hat{T}_{2}$,
\begin{eqnarray}
\left[\hat{J},\hat{G}_a\right]&=&\epsilon_{ab}\hat{G}_b\,, \notag \\
\left[\hat{G}_a,\hat{G}_b\right]&=&-\epsilon_{ab} \hat{S}\,.\label{NW}
\end{eqnarray}
On the other hand, the set of generators $\{G_a,J,S,T_1,T_2,\mathcal{Q}_{\alpha}^{+},\mathcal{Q}_{\alpha}^{-},\mathcal{R}_{\alpha} \}$ satisfies a novel supersymmetric extension of the Nappi-Witten algebra,
\begin{eqnarray}
\left[ J,G_{a}\right] &=&\epsilon _{ab}G_{b}\,,\qquad \qquad \ \ \ \, \quad \left[
G_{a},G_{b}\right] =-\epsilon _{ab}S\,,  \notag \\
\left[ J,\mathcal{Q}_{\alpha }^{\pm }\right] &=&-\frac{1}{2}\left( \gamma _{0}\right)
_{\alpha }^{\,\ \beta }\mathcal{Q}_{\beta }^{\pm }\,,\quad \ \ \ \ \ \ \left[
J,\mathcal{R}_{\alpha }\right] =-\frac{1}{2}\left( \gamma _{0}\right) _{\alpha }^{%
\text{ }\beta }\mathcal{R}_{\beta }\,,  \notag \\
\left[ G_{a},\mathcal{Q}_{\alpha }^{+}\right] &=&-\frac{1}{2}\left( \gamma _{a}\right)
_{\alpha }^{\,\ \beta }\mathcal{Q}_{\beta }^{-}\,,\quad \ \ \ \left[ G_{a},\mathcal{Q}_{\alpha
}^{-}\right] =-\frac{1}{2}\left( \gamma _{a}\right) _{\alpha }^{\text{ }%
\beta }\mathcal{R}_{\beta }\,,  \notag \\
\left[ S,\mathcal{Q}_{\alpha }^{+}\right] &=&-\frac{1}{2}\left( \gamma _{0}\right)
_{\alpha }^{\text{ }\beta }\mathcal{R}_{\beta }\,,\quad \ \ \quad \left[
T_{1},\mathcal{Q}_{\alpha }^{\pm}\right] = \pm \frac{1}{2}\left( \gamma _{0}\right) _{\alpha
\beta }\mathcal{Q}_{\beta }^{\pm}\,,  \notag \\
\left[ T_{2},\mathcal{Q}_{\alpha }^{+}%
\right] &=& \frac{1}{2}\left( \gamma _{0}\right) _{\alpha \beta }\mathcal{R}_{\beta } \,,\qquad \quad \,\,\,\, \left[ T_{1},\mathcal{R}_{\alpha }\right] = \frac{1}{2}\left( \gamma _{0}\right)
_{\alpha \beta }\mathcal{R}_{\beta } \,, \notag \\
\left\{ \mathcal{Q}_{\alpha }^{+},\mathcal{Q}_{\beta }^{-}\right\} &=&-\left( \gamma
^{a}C\right) _{\alpha \beta }G_{a}\,,  \notag \\
\left\{ \mathcal{Q}_{\alpha }^{+},\mathcal{Q}_{\beta }^{+}\right\} &=&-\left( \gamma
^{0}C\right) _{\alpha \beta }J-\left( \gamma ^{0}C\right) _{\alpha \beta
}T_{1}\,,\text{ \qquad }  \notag \\
\left\{ \mathcal{Q}_{\alpha }^{-},\mathcal{Q}_{\beta }^{-}\right\} &=&-\left( \gamma
^{0}C\right) _{\alpha \beta }S {+} \left( \gamma ^{0}C\right) _{\alpha \beta
}T_{2}\,,  \notag \\
\left\{ \mathcal{Q}_{\alpha }^{+},\mathcal{R}_{\beta }\right\} &=&-\left( \gamma ^{0}C\right)
_{\alpha \beta }S-\left( \gamma ^{0}C\right) _{\alpha \beta }T_{2}\,.  \label{sNW}
\end{eqnarray}
Furthermore, one can see that the generators $\{G_a^{*},J^{*},S^{*},T_1^{*},T_2^{*},\mathcal{G}_{\alpha}^{+},\mathcal{G}_{\alpha}^{-},\mathcal{W}_{\alpha} \}$ satisfy an additional copy of a Nappi-Witten superalgebra \eqref{sNW}. The present supersymmetric extension of the Nappi-Witten algebra contains additional bosonic generators $\{T_1,T_2 \}$ which are required to assure the Jacobi identities. Furthermore, as in the case of the extended Bargmann superalgebra \cite{Bergshoeff:2016lwr} and the extended Newtonian superalgebra \cite{Ozdemir:2019tby}, the Nappi-Witten superalgebra has three fermionic charges given by $\mathcal{Q}_{\alpha}
^{+}$, $\mathcal{Q}_{\alpha}^{-}$, and $\mathcal{R}_{\alpha}$. Such feature could be useful to establish some Lie algebra expansion between the Nappi-Witten superalgebra \eqref{sNW} and known NR superalgebras, similarly as in the bosonic case \cite{Concha:2019lhn,Penafiel:2019czp}. Interestingly, the EEB superalgebra inherits the same structure than its relativistic version, in which case the AdS-$\mathcal{L}$ superalgebra can be rewritten as the direct sum of the Lorentz algebra and two super-Lorentz.

Let us note that the Nappi-Witten algebra \eqref{NW} endowed with $\mathfrak{u}(1)$ generators along with the Nappi-Witten superalgebra \eqref{sNW} can be rewritten as an extended Newton-Hooke superalgebra \cite{Ozdemir:2019tby}. Indeed, such structure can be obtained by considering the following redefinition of the generators of \eqref{NW} and \eqref{sNW}:
\begin{eqnarray}
\check{G}_{a}&=&G_a-\hat{G}_a\,, \qquad \qquad \ \  \check{P}_a=\frac{1}{\ell}\left(G_a+\hat{G}_a \right)\,, \qquad \qquad \ \  \check{Q}_{\alpha}^{+}=\sqrt{\frac{2}{\ell}} \mathcal{Q}_{\alpha}^{+}\,,\notag \\
\check{J}&=&J+\hat{J}\,, \qquad \qquad \qquad \check{H}=\frac{1}{\ell}\left(J-\hat{J} \right)\,, \qquad \qquad \qquad \check{Q}_{\alpha}^{-}=\sqrt{\frac{2}{\ell}} \mathcal{Q}_{\alpha}^{-}\,,\notag \\
\check{S}&=&S+\hat{S}\,, \qquad \qquad \qquad \check{M}=\frac{1}{\ell}\left(S-\hat{S} \right)\,, \qquad \qquad \qquad \check{R}_{\alpha}=\sqrt{\frac{2}{\ell}} \mathcal{R}_{\alpha}\,,\notag \\
\check{T}_{1}&=&T_1-\hat{T}_1\,, \qquad \qquad \quad \  \check{U}_1=\frac{1}{\ell}\left(T_1+\hat{T}_1 \right)\,, \notag \\
\check{T}_{2}&=&T_2-\hat{T}_2\,, \qquad \qquad \quad \ \check{U}_2=\frac{1}{\ell}\left(T_2+\hat{T}_2 \right)\,. \label{redef1b}
\end{eqnarray}
One can see that the set of generators $\{\check{G}_a,\check{P}_a,\check{J},\check{H},\check{S},\check{M},\check{T}_1,\check{T}_2,\check{U}_1,\check{U}_2,\check{Q}_{\alpha}^{+},\check{Q}_{\alpha}^{-},\check{R}_{\alpha} \}$ satisfies an extended Newton-Hooke superalgebra,
\begin{eqnarray}
\left[\check{J},\check{G}_a \right]&=&\epsilon_{ab}\check{G}_b\,, \qquad \qquad \ \ \left[ \check{G}_a,\check{G}_b \right]=-\epsilon_{ab}\check{S}\,,\qquad \qquad \quad \ \left[\check{J},\check{P}_a\right]=\epsilon_{ab}\check{P}_b\,, \notag\\
\left[\check{H},\check{G}_a \right]&=&\epsilon_{ab}\check{P}_b\,, \qquad \qquad \ \ \ \left[ \check{G}_a,\check{P}_b \right]=-\epsilon_{ab}\check{M}\,,\qquad \quad \quad \ \ \left[\check{H},\check{P}_a\right]=\frac{1}{\ell^{2}}\epsilon_{ab}\check{G}_b\,, \notag\\
\left[\check{P}_a,\check{P}_b \right]&=&-\frac{1}{\ell^{2}}\epsilon_{ab}\check{S}\,, \quad \qquad \ \ \left[ \check{J}, \check{Q}^{\pm}_\alpha \right]=- \frac{1}{2} \left(\gamma_0\right)_\alpha^{\phantom{\alpha} \beta} \check{Q}^{\pm}_\beta \,, \quad \ \left[ \check{H}, \check{Q}^{\pm}_\alpha \right]=- \frac{1}{2 \ell} \left(\gamma_0\right)_\alpha^{\phantom{\alpha} \beta} \check{Q}^{\pm}_\beta \,, \notag\\
\left[ \check{J}, \check{R}_\alpha \right]&=&- \frac{1}{2} \left(\gamma_0\right)_\alpha^{\phantom{\alpha} \beta} \check{R}_\beta \,, \ \ \ \ \left[ \check{H}, \check{R}_\alpha \right]=- \frac{1}{2 \ell} \left(\gamma_0\right)_\alpha^{\phantom{\alpha} \beta} \check{R}_\beta \,, \ \ \, \left[ \check{G}_a, \check{Q}^{+}_\alpha \right]=- \frac{1}{2} \left(\gamma_a\right)_\alpha^{\phantom{\alpha} \beta} \check{Q}^{-}_\beta \,, \notag \\
\left[ \check{P}_a, \check{Q}^{+}_\alpha \right]&=&- \frac{1}{2 \ell} \left(\gamma_a\right)_\alpha^{\phantom{\alpha} \beta} \check{Q}^{-}_\beta \,, \ \left[ \check{G}_a, \check{Q}^{-}_\alpha \right]=- \frac{1}{2} \left(\gamma_a\right)_\alpha^{\phantom{\alpha} \beta} \check{R}_\beta \,, \ \ \ \ \left[ \check{P}_a, \check{Q}^{-}_\alpha \right]=- \frac{1}{2 \ell} \left(\gamma_a\right)_\alpha^{\phantom{\alpha} \beta} \check{R}_\beta \,, \notag \\
\left[ \check{S}, \check{Q}^{+}_\alpha \right]&=&- \frac{1}{2} \left(\gamma_0\right)_\alpha^{\phantom{\alpha} \beta} \check{R}_\beta \,, \quad \, \left[ \check{M}, \check{Q}^{+}_\alpha \right]=- \frac{1}{2 \ell} \left(\gamma_0\right)_\alpha^{\phantom{\alpha} \beta} \check{R}_\beta \,, \ \ \ \left[ \check{T}_1, \check{Q}^{\pm}_\alpha \right]= \pm \frac{1}{2} \left(\gamma_0\right)_{\alpha \beta} \check{Q}^{\pm}_\beta \,, \notag \\
\left[ \check{U}_1, \check{Q}^{\pm}_\alpha \right]&=& \pm \frac{1}{2 \ell} \left(\gamma_0\right)_{\alpha \beta} \check{Q}^{\pm}_\beta \,, \ \ \left[ \check{T}_2, \check{Q}^{+}_\alpha \right]= \frac{1}{2} \left(\gamma_0\right)_{\alpha \beta} \check{R}_\beta \,, \quad \ \ \ \, \left[ \check{U}_2, \check{Q}^{+}_\alpha \right]= \frac{1}{2 \ell} \left(\gamma_0\right)_{\alpha \beta} \check{R}_\beta \,, \notag \\
\left[ \check{T}_1, \check{R}_\alpha \right]&=& \frac{1}{2} \left(\gamma_0\right)_{\alpha \beta} \check{R}_\beta \,, \quad \quad \left[ \check{U}_1, \check{R}_\alpha \right]= \frac{1}{2 \ell} \left(\gamma_0\right)_{\alpha \beta} \check{R}_\beta \,, \notag \\
\left\{ \check{Q}_{\alpha }^{+},\check{Q}_{\beta }^{-} \right\} &=&- \frac{1}{\ell} \left( \gamma ^{a}C\right)
_{\alpha \beta } \check{G}_a - \left( \gamma ^{a}C\right)
_{\alpha \beta } \check{P}_a \,, \notag \\
\left\{ \check{Q}_{\alpha }^{+},\check{Q}_{\beta }^{+}\right\} &=&- \frac{1}{\ell} \left( \gamma ^{0}C\right)
_{\alpha \beta } \check{J} -\left( \gamma ^{0}C\right) _{\alpha \beta }\check{H} - \frac{1}{\ell} \left( \gamma ^{0}C\right)_{\alpha \beta } \check{T}_1  -\left( \gamma ^{0}C\right) _{\alpha \beta }\check{U}_1 \,, \notag \\
\left\{ \check{Q}_{\alpha }^{-},\check{Q}_{\beta }^{-} \right\} &=&- \frac{1}{\ell} \left( \gamma ^{0}C\right)
_{\alpha \beta } \check{S} - \left( \gamma ^{0}C\right)
_{\alpha \beta } \check{M} + \frac{1}{\ell} \left( \gamma ^{0}C\right) _{\alpha \beta } \check{T}_{2} + \left( \gamma ^{0}C\right) _{\alpha \beta } \check{U}_{2} \,, \notag \\
\left\{ \check{Q}_{\alpha }^{+},\check{R}_{\beta }\right\} &=&- \frac{1}{\ell} \left( \gamma ^{0}C\right)
_{\alpha \beta } \check{S} - \left( \gamma ^{0}C\right)
_{\alpha \beta } \check{M} - \frac{1}{\ell} \left( \gamma ^{0}C\right) _{\alpha \beta } \check{T}_{2} - \left( \gamma ^{0}C\right) _{\alpha \beta } \check{U}_{2} \,, \label{sENH}
\end{eqnarray}
which involves two more generators ($\check{T}_1$ and $\check{T}_2$, which act non-trivially on the fermionic generators $\check{Q}^{\pm}$ and $\check{R}$) with respect to the extended Newton-Hooke superalgebra of \cite{Ozdemir:2019tby}. As we can see, in \eqref{sENH} both $\check{U}_1$ and $\check{U}_2$ act non-trivially on the fermionic generators $\check{Q}^{\pm}$ and $\check{R}$ in the presence of the length parameter $\ell$. In particular, both of these NR generators become central in the flat limit $\ell \rightarrow \infty$ which reproduces a central extension of the extended Bargmann superalgebra \cite{Bergshoeff:2016lwr} endowed with two additional generators $\check{T}_1$ and $\check{T}_2$. Let us note that the extended Newton-Hooke superalgebra \eqref{sENH} can be seen as the NR counterpart of the $\mathcal{N}=2$ AdS superalgebra \cite{Howe:1995zm},
\begin{eqnarray}
\left[ J_A,J_B \right]&=&\epsilon_{ABC}J^{C}\,, \qquad \qquad \quad \left[ J_A,P_B \right]=\epsilon_{ABC}P^{C}\,, \notag \\
\left[ P_A,P_B \right]&=&\frac{1}{\ell^{2}}\epsilon_{ABC}J^{C}\,, \qquad \quad \ \ \left[ J_A,Q_{\alpha}^{i} \right]=-\frac{1}{2} \left( \gamma_{A} \right)_{\alpha}^{\ \beta} Q_{\alpha}^{i} \,, \notag \\
\left[ P_A,Q_{\alpha}^{i} \right]&=&-\frac{1}{2\ell} \left( \gamma_{A} \right)_{\alpha}^{\ \beta} Q_{\alpha}^{i} \,, \qquad \ \, \left[ T,Q_{\alpha}^{i} \right]=\frac{1}{2}\epsilon^{ij} Q_{\alpha}^{i} \, \notag \\ 
\left[U,Q_{\alpha}^{i}\right]&=&\frac{1}{2\ell^{2}}\epsilon^{ij}Q_{\alpha}^{i}\,, \notag \\
\{ Q_{\alpha}^{i},Q_{\beta}^{j}\}&=&-\frac{\delta_{ij}}{\ell}\left( \gamma^{A}C\right)_{\alpha\beta}J_A -\delta_{ij} \left( \gamma^{A}C\right)_{\alpha\beta} P_A - C_{\alpha\beta}\epsilon^{ij} \left(\frac{1}{\ell}T+U\right)\,. \label{sADS}
\end{eqnarray}
Indeed, the extended Newton-Hooke superalgebra \eqref{sENH} can alternatively be obtained after applying a resonant $S_{E}^{(2)}$-expansion of the $\mathcal{N}=2$ AdS superlalgebra and performing a $0_s$-reduction.

Thus, the EEB superalgebra can alternatively be rewritten as the direct sum of the extended Newton-Hooke superalgebra \eqref{sENH} and the super Nappi-Witten one \eqref{sNW}. Such feature is inherited from its relativistic version, in which case the $\mathcal{N}=2$ AdS-$\mathcal{L}$ superalgebra can be written as the direct sum of the super AdS algebra and the super-Lorentz one.

\subsection{Non-relativistic Chern-Simons supergravity action}

We now construct a NR CS supergravity action based on the EEB superalgebra given by \eqref{EEB1}, \eqref{SEEBa}, \eqref{SEEBb}, and \eqref{SEEBc}. Although the superalgebra seems simpler written as copies of the Nappi-Witten (super)algebra, the motivation to consider the EEB structure given by \eqref{EEB1}, \eqref{SEEBa}, \eqref{SEEBb}, and \eqref{SEEBc} is twofold. First, as we shall see, it directly offers us an alternative NR supergravity theory in presence of a cosmological constant. Second, it allows to establish a flat limit leading in a manifest way to the Maxwellian version of the extended Bargmann supergravity \cite{Concha:2019mxx}. 

As we have already mentioned, a crucial ingredient to construct a CS action is the invariant tensor. Interestingly, the same semigroup allowing to obtain the new NR superalgebra can be used to find the non-vanishing components of the EEB invariant tensor (for further details see Appendix \ref{appa}). Indeed, the invariant tensor for the EEB superalgebra can be obtained in terms of the invariant tensor for the $\mathcal{N}=2$ AdS-$\mathcal{L}$ superalgabra given by \eqref{relinvt}. One can then show that the non-vanishing components of the invariant tensor
for the EEB superalgebra are given by (\ref{invt1}) along with
\begin{eqnarray}
\left\langle \tilde{Y}_{1}\tilde{Y}_{2}\right\rangle &=&\tilde{\alpha}_{0}\,, \qquad \qquad \ \ \left\langle \tilde{Y}_{1}\tilde{U}_{2}\right\rangle =\tilde{\alpha}_{1}\,, \qquad \qquad \ \  \left\langle \tilde{U}_{1}\tilde{Y}_{2}\right\rangle = \tilde{\alpha}_{1} \,,  \notag \\
\left\langle \tilde{Y}_{1}\tilde{B}_{2}\right\rangle &=&\tilde{\alpha}_{2}\,, \qquad \qquad \ \ \left\langle \tilde{U}_{1}\tilde{U}_{2}\right\rangle = \tilde{\alpha}_{2}\,, \qquad \qquad \ \ \left\langle \tilde{B}_{1}\tilde{Y}_{2}\right\rangle = \tilde{\alpha}_{2}\,,\notag \\
\left\langle \tilde{U}_{1}\tilde{B}_{2}\right\rangle &=&\frac{\tilde{\alpha}%
_{1}}{\ell ^{2}}\,, \qquad \qquad \ \left\langle \tilde{B}_{1}\tilde{U}_{2}\right\rangle = \frac{\tilde{\alpha}_{1}}{\ell^{2}}\,, \qquad \qquad \ \left\langle \tilde{B}_{1}\tilde{B}_{2}\right\rangle =\frac{\tilde{\alpha}%
_{2}}{\ell ^{2}}\,,  \notag \\
\left\langle \tilde{Q}_{\alpha }^{-}\tilde{Q}_{\beta }^{-}\right\rangle &=&2%
\tilde{\alpha}_{1}C_{\alpha \beta }\,, \quad \quad \left\langle \tilde{Q}_{\alpha }^{+}%
\tilde{R}_{\beta }\right\rangle = 2\tilde{\alpha}_{1}C_{\alpha\beta}\,, \quad \quad \left\langle \tilde{Q}_{\alpha }^{-}\tilde{\Sigma}_{\beta }^{-}\right\rangle
=2\tilde{\alpha}_{2}C_{\alpha \beta } \,, \notag \\
\left\langle \tilde{\Sigma}%
_{\alpha }^{+}\tilde{R}_{\beta }\right\rangle &=& 2\tilde{\alpha}_{2}C_{\alpha\beta}\,, \quad \quad \left\langle \tilde{Q}%
_{\alpha }^{+}\tilde{W}_{\beta }\right\rangle =2\tilde{\alpha}_{2}C_{\alpha\beta}\,, \quad \quad
\left\langle \tilde{\Sigma}_{\alpha }^{+}\tilde{W}_{\beta }\right\rangle =%
\frac{2\tilde{\alpha}_{1}}{\ell ^{2}}C_{\alpha \beta }\,, \notag \\
\left\langle \tilde{\Sigma}
_{\alpha }^{-}\tilde{\Sigma}_{\beta }^{-}\right\rangle &=& \frac{2\tilde{\alpha}_{1}}{\ell^{2}}C_{\alpha\beta}\,,\label{invt2p}
\end{eqnarray}
where the NR generators are related to the relativistic ones through the semigroup elements as in \eqref{sexp1}. Here, the NR constants $\tilde{\alpha}_{i}$ are related to the relativistic parameters appearing in \eqref{relinvt} as
\begin{equation}
\tilde{\alpha}_{0}=\lambda_2\alpha_0\,, \qquad \qquad \tilde{\alpha}_{1}=\lambda_2\alpha_1\,, \qquad \qquad \tilde{\alpha}_{2}=\lambda_2\alpha_2\,.\label{sexp2a}
\end{equation}
On the other hand, the gauge connection one-form $A$ for the EEB superalgebra reads
\begin{eqnarray}
A &=&\omega \tilde{J}+\omega ^{a}\tilde{G}_{a}+\tau \tilde{H}+e^a
\tilde{P}_{a}+k\tilde{Z}+k^{a}\tilde{Z}_{a}+m\tilde{M}+s\tilde{S}+t\tilde{T}+y_{1}\tilde{Y_{1}}+y_{2}\tilde{Y}_{2}+b_{1}\tilde{B}_{1}
\notag \\
&&+b_{2}\tilde{B}%
_{2}+u_{1}\tilde{U}_{1}+u_{2}\tilde{U}_{2}+\bar{\psi }^{+}\tilde{Q}^{+}+\bar{\psi }^{-}\tilde{Q}^{-}+\bar{\xi }^{+}\tilde{\Sigma}%
^{+}+\bar{\xi }^{-}\tilde{\Sigma}^{-}+\bar{\rho }\tilde{R}+\bar{\chi }\tilde{W} \,.
\label{oneform2}
\end{eqnarray}
The corresponding curvature two-form $F=dA+A\wedge
A=dA+\frac{1}{2}\left[ A,A\right] $ written
in terms of the generators is given by
\begin{eqnarray}
F &=&F\left( \omega \right) \tilde{J}+F^{a}\left( \omega ^{b}\right)
\tilde{G}_{a}+F\left( \tau \right) \tilde{H}+F^{a}\left( e^{b}\right) \tilde{%
P}_{a}+F\left( k\right) \tilde{Z}+F^{a}\left( k^{b}\right) \tilde{Z}%
_{a}+F\left( m\right) \tilde{M}  \notag \\
&&+F\left( s\right) \tilde{S}+F\left( t\right) \tilde{T}+F\left(
y_{1}\right) \tilde{Y_{1}}+F\left( y_{2}\right) \tilde{Y}_{2}+F\left(
b_{1}\right) \tilde{B}_{1}+F\left( b_{2}\right) \tilde{B}_{2}+F\left(
u_{1}\right) \tilde{U}_{1}  \notag \\
&&+F\left( u_{2}\right) \tilde{U}_{2}+\nabla \bar{\psi }^{+}\tilde{Q}^{+}+\nabla \bar{\psi }^{-}\tilde{Q}^{-}+\nabla \bar{\xi }^{+}\tilde{\Sigma}^{+}+\nabla \bar{\xi }^{-}\tilde{\Sigma}^{-}+\nabla \bar{\rho
}\tilde{R}+\nabla \bar{\chi }\tilde{W}\,.  \label{F2c}
\end{eqnarray}%
In particular, the bosonic curvature two-forms are given by%
\begin{eqnarray}
F\left( \omega \right) &=&R\left( \omega \right) \,,  \notag \\
F^{a}\left( \omega ^{b}\right) &=&R^{a}\left( \omega ^{b}\right) \,,  \notag
\\
F\left( \tau \right) &=&R\left( \tau \right) +\frac{1}{2}\bar{\psi}%
^{+}\gamma ^{0}\psi ^{+}+\frac{1}{2\ell ^{2}}\bar{\xi}^{+}\gamma ^{0}\xi
^{+}\,,  \notag \\
F^{a}\left( e^{b}\right) &=&R^{a}\left( e^{b}\right) +\bar{\psi}^{+}\gamma
^{a}\psi ^{-}+\frac{1}{\ell ^{2}}\bar{\xi}^{+}\gamma ^{a}\xi ^{-}\,,  \notag
\\
F\left( k\right) &=&R\left( k\right) +\bar{\psi}^{+}\gamma ^{0}\xi ^{+}\,,
\notag \\
F^{a}\left( k^{b}\right) &=&R^{a}\left( k^{b}\right) +\bar{\psi}^{+}\gamma
^{a}\xi ^{-}+\bar{\psi}^{-}\gamma ^{a}\xi ^{+}\,,  \notag \\
F\left( m\right) &=&R\left( m\right) +\frac{1}{2}\bar{\psi}^{-}\gamma
^{0}\psi ^{-}+\bar{\psi}^{+}\gamma ^{0}\rho +\frac{1}{2\ell ^{2}}\bar{\xi}%
^{-}\gamma ^{0}\xi ^{-}+\frac{1}{\ell ^{2}}\bar{\xi}^{+}\gamma ^{0}\chi \,,
\notag \\
F\left( s\right) &=&R\left( s\right) \,,  \notag \\
F\left( t\right) &=&R\left( t\right) +\bar{\psi}^{-}\gamma ^{0}\xi ^{-}+\bar{%
\psi}^{+}\gamma ^{0}\chi +\bar{\xi}^{+}\gamma ^{0}\rho \,,  \label{boscurvSuperEEBp1}
\end{eqnarray}%
where $R\left( \omega \right) $, $R^{a}\left( \omega ^{b}\right) $, $R\left(
\tau \right) $, $R^{a}\left( e^{b}\right) $, $R\left( k\right) $, $%
R^{a}\left( k^{b}\right) $, $R\left( m\right) $, $R\left( s\right) $, and $%
R\left( t\right) $ are the EEB curvatures defined in (\ref{curvEEB}),
together with
\begin{eqnarray}
F\left( y_{1}\right) &=&dy_{1}\,,  \notag \\
F\left( y_{2}\right) &=&dy_{2}\,,  \notag \\
F\left( b_{1}\right) &=&db_{1}+\bar{\psi}^{+}\gamma ^{0}\xi ^{+}\,,  \notag
\\
F\left( b_{2}\right) &=&db_{2}-\bar{\psi}^{-}\gamma ^{0}\xi ^{-}+\bar{\psi}%
^{+}\gamma ^{0}\chi +\bar{\xi}^{+}\gamma ^{0}\rho \,,
\notag \\
F\left( u_{1}\right) &=&du_{1}+\frac{1}{2}\bar{\psi}^{+}\gamma ^{0}\psi ^{+}+%
\frac{1}{2\ell ^{2}}\bar{\xi}^{+}\gamma ^{0}\xi ^{+}\,,  \notag \\
F\left( u_{2}\right) &=&du_{2}-\frac{1}{2}\bar{\psi}^{-}\gamma ^{0}\psi ^{-}+%
\bar{\psi}^{+}\gamma ^{0}\rho -\frac{1}{2\ell ^{2}}\bar{\xi}^{-}\gamma
^{0}\xi ^{-}+\frac{1}{\ell ^{2}}\bar{\xi}^{+}\gamma ^{0}\chi \,.  \label{boscurvSuperEEBp2}
\end{eqnarray}%
Let us observe that, if we restrict ourselves to the purely bosonic sector and excluding the extra bosonic gauge field $\{y_1,y_2,u_1,u_2,b_1,b_2\}$,
we properly recover the two-form curvatures associated with the EEB algebra
introduced in \cite{Concha:2019lhn}. On the other hand, the covariant
derivatives of the spinor $1$-form fields read
\begin{eqnarray}
\nabla \psi ^{+} &=&d\psi ^{+}+\frac{1}{2}\omega \gamma _{0}\psi ^{+}-\frac{1%
}{2}y_{1}\gamma _{0}\psi ^{+}+\frac{1}{2\ell ^{2}}\tau \gamma _{0}\xi ^{+}+%
\frac{1}{2\ell ^{2}}k\gamma _{0}\psi ^{+}-\frac{1}{2\ell ^{2}}u_{1}\gamma
_{0}\xi ^{+}-\frac{1}{2\ell ^{2}}b_{1}\gamma _{0}\psi ^{+}\,,  \notag \\
\nabla \psi ^{-} &=&d\psi ^{-}+\frac{1}{2}\omega \gamma _{0}\psi ^{-}+\frac{1%
}{2}\omega ^{a}\gamma _{a}\psi ^{+}+\frac{1}{2}y_{1}\gamma _{0}\psi ^{-}+%
\frac{1}{2\ell ^{2}}\tau \gamma _{0}\xi ^{-}+\frac{1}{2\ell ^{2}}e^{a}\gamma
_{a}\xi ^{+}+\frac{1}{2\ell ^{2}}k^{a}\gamma _{a}\psi ^{+}  \notag \\
&&+\frac{1}{2\ell ^{2}}k\gamma _{0}\psi ^{-}+\frac{1}{2\ell ^{2}}u_{1}\gamma
_{0}\xi ^{-}+\frac{1}{2\ell ^{2}}b_{1}\gamma _{0}\psi ^{-}\,,  \notag \\
\nabla \xi ^{+} &=&d\xi ^{+}+\frac{1}{2}\omega \gamma _{0}\xi ^{+}+\frac{1}{2%
}\tau \gamma _{0}\psi ^{+}-\frac{1}{2}y_{1}\gamma _{0}\xi ^{+}-\frac{1}{2}%
u_{1}\gamma _{0}\psi ^{+}+\frac{1}{2\ell ^{2}}k\gamma _{0}\xi ^{+}-\frac{1}{%
2\ell ^{2}}b_{1}\gamma _{0}\xi ^{+}\,,  \notag \\
\nabla \xi ^{-} &=&d\xi ^{-}+\frac{1}{2}\omega \gamma _{0}\xi ^{-}+\frac{1}{2%
}\tau \gamma _{0}\psi ^{-}+\frac{1}{2}e^{a}\gamma _{a}\psi ^{+}+\frac{1}{2}%
\omega ^{a}\gamma _{a}\xi ^{+}+\frac{1}{2}y_{1}\gamma _{0}\xi ^{-}+\frac{1}{2%
}u_{1}\gamma _{0}\psi ^{-}  \notag \\
&&+\frac{1}{2\ell ^{2}}k^{a}\gamma _{a}\xi ^{+}+\frac{1}{2\ell ^{2}}k\gamma
_{0}\xi ^{-}+\frac{1}{2\ell ^{2}}b_{1}\gamma _{0}\xi ^{-}\,,
\notag \\
\nabla \rho &=&d\rho +\frac{1}{2}\omega \gamma _{0}\rho +\frac{1}{2}\omega
^{a}\gamma _{a}\psi ^{-}+\frac{1}{2}s\gamma _{0}\psi ^{+}-\frac{1}{2}%
y_{2}\gamma _{0}\psi ^{+}-\frac{1}{2}y_{1}\gamma _{0}\rho +\frac{1}{2\ell
^{2}}e^{a}\gamma _{a}\xi ^{-}  \notag \\
&&+\frac{1}{2\ell ^{2}}k^{a}\gamma _{a}\psi ^{-}+\frac{1}{2\ell ^{2}}m\gamma
_{0}\xi ^{+}+\frac{1}{2\ell ^{2}}\tau \gamma _{0}\chi +\frac{1}{2\ell ^{2}}%
t\gamma _{0}\psi ^{+}+\frac{1}{2\ell ^{2}}k\gamma _{0}\rho  \notag \\
&&-\frac{1}{2\ell ^{2}}u_{1}\gamma _{0}\chi -\frac{1}{2\ell ^{2}}u_{2}\gamma
_{0}\xi ^{+}-\frac{1}{2\ell ^{2}}b_{1}\gamma _{0}\rho -\frac{1}{2\ell ^{2}}%
b_{2}\gamma _{0}\psi ^{+}\,,  \notag \\
\nabla \chi &=&d\chi +\frac{1}{2}\omega \gamma _{0}\chi +\frac{1}{2}\omega
^{a}\gamma _{a}\xi ^{-}+\frac{1}{2}e^{a}\gamma _{a}\psi ^{-}+\frac{1}{2}\tau
\gamma _{0}\rho +\frac{1}{2}s\gamma _{0}\xi ^{+}+\frac{1}{2}m\gamma _{0}\psi
^{+}  \notag \\
&&-\frac{1}{2}y_{2}\gamma _{0}\xi ^{+}-\frac{1}{2}y_{1}\gamma _{0}\chi -%
\frac{1}{2}u_{2}\gamma _{0}\psi ^{+}-\frac{1}{2}u_{1}\gamma _{0}\rho +\frac{1%
}{2\ell ^{2}}k^{a}\gamma _{a}\xi ^{-}+\frac{1}{2\ell ^{2}}t\gamma _{0}\xi
^{+}+\frac{1}{2\ell ^{2}}k\gamma _{0}\chi  \notag \\
&&-\frac{1}{2\ell ^{2}}b_{1}\gamma _{0}\chi -\frac{1}{2\ell ^{2}}b_{2}\gamma
_{0}\xi ^{+}\,.  \label{fermcurvSuperEEB}
\end{eqnarray}%
Notice that taking the $\ell \rightarrow \infty $ limit of the curvature
two-forms (\ref{boscurvSuperEEBp1}), (\ref{boscurvSuperEEBp2}), and (\ref%
{fermcurvSuperEEB}) one recovers the curvatures associated with the MEB
superalgebra of \cite{Concha:2019mxx}.

A CS supergravity action based on the EEB superalgebra given by \eqref{EEB1}, \eqref{SEEBa}, \eqref{SEEBb}, and \eqref{SEEBc} can be constructed by combining the non-zero components of the
invariant tensor \eqref{invt1}  and \eqref{invt2p} with
the gauge connection one-form $A$ \eqref{oneform2}. The NR CS supergravity action reads, up
to boundary terms, as follows:
\begin{eqnarray}
I_{\text{NR}}&=&\tilde{\alpha}_{0}I_{0}+\tilde{\alpha}_{1}I_{1}+\tilde{\alpha}_{2}I_{2}\,,\label{CS2}
\end{eqnarray}
where
\begin{eqnarray}
I_0 &=&\int  \omega _{a}R^{a}\left(\omega ^{b}\right)-2sR\left(
\omega \right) +2y_{1}dy_{2}\,, \notag \\
I_1&=&\int 
e_{a}R^{a}\left(\omega ^{b}\right) + \omega_{a}R^{a}\left(e^{b}\right)-2mR(\omega )-2\tau ds +\frac{1}{\ell ^{2}}e_{a} {R}^{a}\left(k^{b}\right)+ \frac{1}{\ell ^{2}}k_{a} {R}^{a}\left(e^{b}\right) \notag
\\
&&  -\frac{2}{%
\ell ^{2}}\tau dt -\frac{2}{\ell ^{2}}mR(k)+ 2y_{1}du_{2} + 2u_{1}dy_{2}+\frac{2}{\ell ^{2}}u_{1}db_{2}+\frac{2}{\ell ^{2}}b_{1}du_{2}-2\bar{\psi}^{+}\nabla \rho  \notag \\
&& - 2\bar{\rho}\nabla \psi ^{+}-2\bar{\psi}^{-}\nabla \psi ^{-}-\frac{2%
}{\ell ^{2}}\bar{\xi}^{+}\nabla \chi -\frac{2}{\ell ^{2}}\bar{\chi}\nabla
\xi ^{+} - \frac{2}{\ell ^{2}}\bar{\xi}^- \nabla
\xi ^{-}\,, \notag \\
I_2&=&\int  e_{a}R^{a}\left( e^{b}\right)
+k_{a}R^{a}\left( \omega ^{b}\right) +\omega _{a}R^{a}\left( k^{b}\right) + \frac{1}{\ell ^{2}}k_{a}R^{a}\left(k^{b}\right)-2sR\left( k\right) -2mR\left(
\tau \right)   \notag \\
&&- 2tR\left( \omega \right) -\frac{2}{\ell ^{2}}
tR(k) + 2y_{1}db_{2}+2u_{1}du_{2}+2y_{2}db_{1} + \frac{2}{\ell ^{2}}b_{1}db_{2}-2\bar{\psi}^{-}\nabla \xi ^{-}   \notag \\
&&- 2%
\bar{\xi}^{-}\nabla \psi ^{-} -  2\bar{\psi}^{+}\nabla \chi -2\bar{\chi}\nabla
\psi ^{+}-2\bar{\xi}^{+}\nabla \rho -2\bar{\rho}\nabla \xi ^{+} \,.
\label{CS2b}
\end{eqnarray}
The CS action (\ref{CS2}) obtained here describes the so-called Enlarged
Extended Bargmann supergravity theory. Let us note that the NR CS
supergravity action (\ref{CS2}) contains three independent sectors
proportional to $\tilde{\alpha}_{0}$, $\tilde{\alpha}_{1}$, and $\tilde{\alpha}_{2}$. In particular, $I_0$ corresponds to the CS action for the NR exotic gravity coupled to the extra gauge fields $y_1$ and $y_2$. The CS actions $I_1$ and $I_2$ describe the EEB CS supergravity in presence of the cosmological constant and the gauge field $k_a$. We observe that taking the flat limit $\ell \rightarrow \infty$ of \eqref{CS2} we recover the CS MEB supergravity action of \cite{Concha:2019mxx}.\footnote{This also fixes a sign misprint appearing in \cite{Concha:2019mxx}.} In particular, the vanishing cosmological constant limit applied to the CS action $I_1$ reproduces the extended Bargmann supergravity CS action coupled to the additional gauge fields $y_1$, $y_2$, $u_1$, and $u_2$. The new NR CS supergravity action \eqref{CS2} generalizes the extended Newton-Hooke supergravity theory \cite{Ozdemir:2019tby} in the sense that it not only reproduces the extended Bargmann supergravity in the flat limit, but also the Maxwellian version proportional to $I_2$.

It is important to emphasize that in order to obtain the CS supergravity action through the $S$-expansion method we have restricted ourselves exclusively to the CS terms related to the $\lambda_2$ element of the semigroup $S_{E}^{(2)}$. Indeed, all the NR parameter $\tilde{\alpha}_0$, $\tilde{\alpha}_{1}$, and $\tilde{\alpha}_{2}$ are defined as in \eqref{sexp2a} in terms of the relativistic parameters and the $\lambda_2$ element. On the other hand, it is possible to obtain diverse exotic-like contributions to the CS action which are proportional to the $\lambda_0$ element. Nevertheless, since we are interested in the supersymmetric extension of the EEB gravity theory \cite{Concha:2019lhn}, we shall omit such exotic-like terms which would imply that the bosonic sector is no more EEB gravity. It would be interesting to study the Physical implications of such additional contributions.

Let us note that the CS supergravity action \eqref{CS2} can alternatively be obtained by expanding directly the relativistic $\mathcal{N}=2$ AdS-$\mathcal{L}$ CS supergravity action \cite{Concha:2019icz},
\begin{eqnarray}
I_{\text{R}}&=&\int \alpha_0\left[\omega_A d\omega^{A}+\frac{1}{3}\epsilon_{ABC}\omega^{A}\omega^{B}\omega^{C} +\mathtt{t}d\mathtt{t} \right] +\alpha_{1}\left[2e_{A}R^{A}+\frac{1}{3\ell^{2}}\epsilon_{ABC}e^{A}e^{B}e^{C}+\frac{2}{\ell^{2}}e_{A}F^{A} \right. \notag \\
&&\left.-\bar{\hat{\psi}}^{i}\nabla\hat{\psi}^{i}-\frac{1}{\ell^{2}}\bar{\hat{\xi}}^{i}\nabla\hat{\xi}^{i}+\mathtt{u}d\mathtt{t}+\mathtt{u}d\mathtt{b} \right]+\alpha_2 \left[ \right.2\sigma_AR^{A}+e_AT^{A}+\frac{2}{\ell^{2}}e_AF^{A}+\frac{1}{\ell^{2}}\epsilon_{ABC}e^{A}\sigma^{B}e^{C}\notag \\
&&\left. -\bar{\hat{\psi}}^{i}\nabla\hat{\xi}^{i}-\bar{\hat{\xi}}^{i}\nabla\hat{\psi}^{i}+\mathtt{b}d\mathtt{t}+\mathtt{u}d\mathtt{u}+\frac{1}{\ell^{2}}\mathtt{b}d\mathtt{b}\right]\,,\label{N2ADSLCS}
\end{eqnarray}
where
\begin{eqnarray}
R^{A}&=&d\omega^{A}+\frac{1}{2}\epsilon^{ABC}\omega_{B}\omega_{C}\,,\notag\\
T^{A}&=&de^{A}+\epsilon^{ABC}\omega_{B}e_{C}\,, \notag \\
F^{A}&=&d\sigma^{A}+\epsilon^{ABC}\omega_{B}\sigma_{C}+\frac{1}{2\ell^{2}}\epsilon^{ABC}\sigma_{B}\sigma_{C}\,,\notag \\
\nabla\hat{\psi}^{i}&=&d\hat{\psi}^{i}+\frac{1}{2}\omega^{A}\gamma_{A}\hat{\psi}^{i}+\mathtt{t}\epsilon^{ij}\hat{\psi}^{j}+\frac{1}{\ell^{2}}\mathtt{b}\epsilon^{ij}\hat{\psi}^{j}+\frac{1}{\ell^{2}}\mathtt{u}\epsilon^{ij}\hat{\xi}^{j}\,,\notag\\
\nabla\hat{\xi}^{i}&=&d\hat{\xi}^{i}+\frac{1}{2}\omega^{A}\gamma_{A}\hat{\xi}^{i}+\frac{1}{2}e^{A}\gamma_{A}\hat{\psi}^{i}+\mathtt{t}\epsilon^{ij}\hat{\xi}^{j}+\mathtt{u}\epsilon^{ij}\hat{\psi}^{j}+\frac{1}{\ell^{2}}\mathtt{b}\epsilon^{ij}\hat{\xi}^{j}\,. \label{relcurv}
\end{eqnarray}
Indeed, we can express the NR gauge fields in terms of the relativistic ones and the semigroup elements as
\begin{eqnarray}
\omega&=&\lambda_0 \omega_0\,, \qquad  \ \, s=\lambda_2 \omega_0\,, \qquad \omega_a=\lambda_1 \omega_a\,, \notag \\
\tau&=&\lambda_0 e_0\,, \qquad \ m=\lambda_2 e_0\,, \qquad \ e_a=\lambda_1 e_a \,, \notag \\
k&=&\lambda_0 \sigma_0\,, \qquad \ \ \, t=\lambda_2 \sigma_0\,, \qquad  \, k_a=\lambda_1 \sigma_a\,, \notag \\
\psi_{\alpha}^{+}&=&\lambda_0 \hat{\psi}_{\alpha}^{+}\,, \quad \ \  \rho_{\alpha}=\lambda_2 \hat{\psi}_{\alpha}^{+}\,, \quad \ \, \psi_{\alpha}^{-}=\lambda_1 \hat{\psi}_{\alpha}^{-}\,, \notag \\
\xi_{\alpha}^{+}&=&\lambda_0 \hat{\xi}_{\alpha}^{+}\,, \quad \ \ \, \chi_{\alpha}=\lambda_2 \hat{\xi}_{\alpha}^{+}\,, \quad \ \ \, \xi_{\alpha}^{-}=\lambda_1 \hat{\xi}_{\alpha}^{-}\,, \notag \\
y_{1}&=&\lambda_0 \mathtt{t}\,, \qquad \ \ u_1=\lambda_0 \mathtt{u}\,, \qquad \ \   b_1=\lambda_0 \mathtt{b}\,, \notag\\
y_{2}&=&\lambda_2 \mathtt{t}\,, \qquad \ \  u_2=\lambda_2 \mathtt{u}\,, \qquad \ \  b_2=\lambda_2 \mathtt{b}\,,\label{gfredef}
\end{eqnarray}
where we have split the Lorentz index as $A=\lbrace 0,a \rbrace$ with $a=1,2$ and where we have defined
\begin{equation}
    \hat{\psi}_{\alpha}^{\pm}=\frac{1}{\sqrt{2}}\left( \hat{\psi}_{\alpha}^{1}\pm \epsilon_{\alpha\beta}\hat{\psi}_{\beta}^{2}\right)\,, \qquad \qquad
    \hat{\xi}_{\alpha}^{\pm}=\frac{1}{\sqrt{2}}\left( \hat{\xi}_{\alpha}^{1}\pm \epsilon_{\alpha\beta}\hat{\xi}_{\beta}^{2}\right)\,.\label{redeff2}
\end{equation}
Then, the EEB supergravity action is obtained considering the expanded gauge fields \eqref{gfredef}, the expanded parameters \eqref{sexp2a}, the multiplication law \eqref{ml}, and the $0_s$-reduction property. 

As an ending remark let us note that, since the invariant tensor is non-degenerate, the field equations from the NR CS supergravity action (\ref{CS2}) imply the
vanishing of the curvature two-forms (\ref{boscurvSuperEEBp1}), (\ref%
{boscurvSuperEEBp2}), and (\ref{fermcurvSuperEEB}) associated with the EEB
superalgebra. The aforementioned curvatures transform covariantly with
respect to the supersymmetry transformation laws given in Appendix \ref{appb}.


\section{Three-dimensional non-standard enlarged extended Bargmann supergravity}

Let us focus now on an alternative supersymmetric extension of the EEB algebra which we call as non-standard EEB superalgebra. The new structure appears as an $S$-expansion of a different $\mathcal{N}=2$ AdS-$\mathcal{L}$ superalgebra and, unlike the EEB superalgebra presented in the previous section, contains only three fermionic charges. The construction of a CS supergravity action based on the aforementioned superalgebra is also discussed.

\subsection{Non-standard enlarged extended Bargmann superalgebra}

An alternative supersymmetric extension of the EEB algebra can be defined by expanding a $\mathcal{N}=2$ extension of the relativistic non-standard AdS-$\mathcal{L}$ superalgebra \cite{Concha:2018jxx,Concha:2020atg}. The aforesaid $\mathcal{N}=2$ non-standard supersymmetric extension of the AdS-$\mathcal{L}$ algebra is spanned by the set of generators $\{ J_a,P_a,Z_a,\mathcal{U},\mathcal{T},\mathcal{B},Q_{\alpha}^{i} \}$ which satisfy the following (anti-)commutation relations:
\begin{eqnarray}
\left[ J_{A},J_{B}\right] &=&\epsilon _{ABC}J^{C}\,, \qquad \qquad \, \left[ J_{A},P_{B}\right] =\epsilon _{ABC}P^{C}\,, \notag \\
\left[ J_{A},Z_{B}\right] &=&\epsilon _{ABC}Z^{C}\,, \qquad \qquad \left[ P_{A},P_{B}\right] =\epsilon _{ABC}Z^{C}\,, \notag \\
\left[ Z_{A},Z_{B}\right] &=&\frac{1}{\ell ^{2}}\epsilon _{ABC}Z^{C}\,, \qquad \quad \left[ P_{A},Z_{B}\right] =\frac{1}{\ell ^{2}}\epsilon _{ABC}P^{C}\,, \notag \\
\left[J_{A},Q_{\alpha}^{i}\right] &=& -\frac{1}{2} \left( \gamma_{A} \right)_{\alpha}^{\ \beta} Q_{\beta}^{i}\,, \quad \ \, \left[P_{A},Q_{\alpha}^{i}\right] = -\frac{1}{2\ell} \left( \gamma_{A} \right)_{\alpha}^{\ \beta} Q_{\beta}^{i}\,, \notag \\
\left[Z_{A},Q_{\alpha}^{i}\right] &=& -\frac{1}{2\ell^{2}} \left( \gamma_{A} \right)_{\alpha}^{\ \beta} Q_{\beta}^{i}\,, \notag \quad \, \left[\mathcal{T},Q_{\alpha}^{i}\right] = \frac{1}{2} \epsilon^{ij} Q_{\beta}^{j}\,, \notag \\
\left[\mathcal{U},Q_{\alpha}^{i}\right] &=& \frac{1}{2\ell^{2}} \epsilon^{ij} Q_{\beta}^{j}\,, \notag \\
\{ Q_{\alpha}^{i},Q_{\beta}^{j}\} &=&-\frac{\delta_{ij}}{\ell}\left(\gamma^{A}C \right)_{\alpha \beta} P_{A}-\delta_{ij}\left(\gamma^{A}C \right)_{\alpha \beta} Z_{A}-C_{\alpha \beta}\epsilon^{ij}\left(\mathcal{U}+\frac{1}{\ell}\mathcal{B}\right)\,, \label{nsSADSL}
\end{eqnarray}%
where, as usual, $i=1,2$ denotes the number of supercharges and $A,B,\ldots=0,1,2$ are the Lorentz indices. 
Let us note that $\mathcal{B}$ is a central charge, while $\mathcal{U}$ and $\mathcal{T}$ are $\mathfrak{so}(2)$ internal symmetry generators. The presence of the central charge and internal symmetry generators are required to guarantee the non-degeneracy of the invariant tensor. In particular, the non-vanishing components of a non-degenerate invariant tensor of the superalgebra \eqref{nsSADSL} are given by
\begin{eqnarray}
\left\langle J_{A} J_{B}\right\rangle &=&{\alpha}_{0}\eta_{AB}\,, \qquad \qquad \quad \ \
\left\langle J_{A} P_{B}\right\rangle ={\alpha}_{1}\eta_{AB}\,,  \notag \\
\left\langle J_{A} Z_{B}\right\rangle &=&{\alpha}_{2}\eta_{AB}\,, \qquad \qquad \quad \ \
\left\langle P_{A} P_{B}\right\rangle ={\alpha}_{2}\eta_{AB}\,,
\notag \\
\left\langle P_{A} Z_{B}\right\rangle &=&\frac{\alpha_{1}}{\ell^{2}}\eta_{AB}\,, \qquad \qquad \quad \ \, \left\langle Z_{A} Z_{B}\right\rangle =\frac{\alpha_{2}}{\ell^{2}}\eta_{AB}\,, \notag \\
\left\langle \mathcal{T} \mathcal{T} \right\rangle &=& \alpha_{0}\,,\qquad \qquad \qquad \qquad \ \ \left\langle \mathcal{T} \mathcal{U} \right\rangle = \alpha_{2}\,,   \notag\\ 
\left\langle \mathcal{U} \mathcal{U} \right\rangle &=& \frac{1}{\ell^{2}}\alpha_{2}\,, \qquad \qquad \qquad \quad \ \ \notag \left\langle \mathcal{T} \mathcal{B} \right\rangle = \alpha_{1}\,, \\
\left\langle \mathcal{U} \mathcal{B} \right\rangle &=& \frac{1}{\ell^{2}}\alpha_{1}\,, \qquad \qquad \qquad \quad \ \ \, \notag \left\langle \mathcal{B} \mathcal{B} \right\rangle = - \frac{1}{\ell}\alpha_{1}\,, \\
\left\langle Q_{\alpha}^{i} Q_{\beta}^{j} \right\rangle &=& 2\left(\frac{\alpha_1}{\ell} +\alpha_{2} \right) C_{\alpha\beta}\,\delta^{ij}\,.\label{nsinvt}
\end{eqnarray}
Although such superalgebra is well-defined since it allows us to reproduce a proper three-\-dimensional $\mathcal{N}=2$ CS supergravity action in presence of a cosmological constant, its flat limit $\ell\rightarrow\infty$ is problematic. Indeed, the vanishing cosmological constant limit reproduces an $\mathcal{N}=2$ exotic Maxwell supersymmetric CS action due to the behavior of the $P_{A}$ generator which is no more expressed as a bilinear expression of the fermionic generators $Q
^{i}$. Such feature is responsible of the label ``non-standard" in the Maxwell superalgebra \cite{Soroka:2004fj,Lukierski:2010dy}. A dual version of the non-standard super-Maxwell algebra being the supersymmetric extension of the Hietarinta-Maxwell algebra \cite{Bansal:2018qyz,Chernyavsky:2020fqs} could overcome such difficulty by interchanging the role of the $P_A$ and $Z_A$ generators. Nevertheless, such approach will not be considered here.

Here, we shall see that a new and consistent NR superalgebra is obtained by considering an $S$-expansion of the non-standard superalgebra \eqref{nsSADSL}. Let us consider $S_E^{(2)}=\{\lambda_0,\lambda_1,\lambda_2,\lambda_3\}$ as the relevant abelian semigroup whose elements satisfy the multiplication law \eqref{ml} with $\lambda_3=0_S$ being the zero element of the semigroup. Let $S_{E}^{(2)}=S_{0} \cup S_{1}$ be a semigroup decomposition where
\begin{eqnarray}
S_0&=&\{\lambda_0,\lambda_2,\lambda_3\}\,, \notag \\
S_1&=&\{\lambda_1,\lambda_3\} \,. 
\label{decomp2}
\end{eqnarray}
The decomposition \eqref{decomp2} is said to be resonant since it satisfies the same structure than the subspaces $V_0=\{J_0,P_0,Z_0,\mathcal{T},\mathcal{U},\mathcal{B},Q_{\alpha}^{+} \}$ and $V_1=\{J_a,P_a,Z_a,Q_{\alpha}^{-} \}$ of the $\mathcal{N}=2$ superalgebra \eqref{nsSADSL} with $a=1,2$ and where we have defined
\begin{equation}
    Q_{\alpha}^{\pm}=\frac{1}{\sqrt{2}}\left( Q_{\alpha}^{1}\pm \epsilon_{\alpha\beta}Q_{\beta}^{2}\right)\,. \label{redef21}
\end{equation}

A new NR superalgebra is obtained after applying a resonant $S_{E}^{(2)}$-expansion of \eqref{nsSADSL} and performing a $0_S$-reduction. In particular, the expanded generators are related to the $\mathcal{N}=2$ super AdS-$\mathcal{L}$ ones through the semigroup elements as
\begin{eqnarray}
\tilde{J}&=&\lambda_0 J_0\,, \qquad  \ \, \tilde{S}=\lambda_2 J_0\,, \qquad \tilde{G}_a=\lambda_1 J_a\,, \notag \\
\tilde{H}&=&\lambda_0 P_0\,, \qquad \tilde{M}=\lambda_2 P_0\,, \qquad \tilde{P}_a=\lambda_1 P_a \,, \notag \\
\tilde{Z}&=&\lambda_0 Z_0\,, \qquad \ \tilde{T}=\lambda_2 Z_0\,, \qquad \tilde{Z}_a=\lambda_1 Z_a\,, \notag \\
\tilde{Q}_{\alpha}^{+}&=&\lambda_0 Q_{\alpha}^{+}\,, \quad \ \, \tilde{R}_{\alpha}=\lambda_2 Q_{\alpha}^{+}\,, \quad \ \tilde{Q}_{\alpha}^{-}=\lambda_1 Q_{\alpha}^{-}\,, \notag \\
\tilde{Y}_{1}&=&\lambda_0 \mathcal{T}\,, \qquad \  \tilde{U}_1=\lambda_0 \mathcal{U}\,, \qquad \  \tilde{B}_1=\lambda_0 \mathcal{B}\,, \notag\\
\tilde{Y}_{2}&=&\lambda_2 \mathcal{T}\,, \qquad \  \tilde{U}_2=\lambda_2 \mathcal{U}\,, \qquad \  \tilde{B}_2=\lambda_2 \mathcal{B}\,.\label{sexp2}
\end{eqnarray}
The commutators of the expanded NR superalgebra are given by the bosonic EEB algebra \eqref{EEB1} along with the following commutation relations:
\begin{eqnarray}
\left[ \tilde{J},\tilde{Q}_{\alpha }^{\pm }\right] &=&-\frac{1}{2}\left(
\gamma _{0}\right) _{\alpha }^{\text{ }\beta }\tilde{Q}_{\beta }^{\pm
}\,,\qquad \left[ \tilde{J},\tilde{R}_{\alpha }\right] =-%
\frac{1}{2}\left( \gamma _{0}\right) _{\alpha }^{\text{ }\beta }\tilde{R%
}_{\beta }\,,   \qquad \left[ \tilde{H},\tilde{Q}_{\alpha }^{\pm}\right] =-\frac{1}{2\ell}\left(
\gamma _{0}\right) _{\alpha }^{\text{ }\beta }\tilde{Q}_{\beta }^{\pm
}\,,  \notag\\
\left[ \tilde{H},\tilde{R}_{\alpha }\right] &=&-\frac{1}{2\ell}\left(
\gamma _{0}\right) _{\alpha }^{\text{ }\beta }\tilde{R}_{\beta }\,, \quad \ \, \left[ \tilde{Z},\tilde{Q}_{\alpha }^{\pm}\right] =-\frac{1}{2\ell^{2}}\left(
\gamma _{0}\right) _{\alpha }^{\text{ }\beta }\tilde{Q}_{\beta }^{\pm}\,,  \quad \ 
\left[ \tilde{Z},\tilde{R}_{\alpha }\right] =-\frac{1}{2\ell^{2}}\left(
\gamma _{0}\right) _{\alpha }^{\text{ }\beta }\tilde{R}_{\beta }\,, \notag \\
\left[ \tilde{S},\tilde{Q}_{\alpha }^{+}\right] &=&-\frac{1}{2}\left(
\gamma _{0}\right) _{\alpha }^{\text{ }\beta }\tilde{R}_{\beta }\,, \quad \ \
\left[ \tilde{M},\tilde{Q}_{\alpha }^{+}\right] =-\frac{1}{2\ell}\left(
\gamma _{0}\right) _{\alpha }^{\text{ }\beta }\tilde{R}_{\beta }\,, \quad \ \ \, \left[ \tilde{T},\tilde{Q}_{\alpha }^{+}\right] =-\frac{1}{2\ell^{2}}\left(
\gamma _{0}\right) _{\alpha }^{\text{ }\beta }\tilde{R}_{\beta }\,, \notag \\
\left[ \tilde{G}_{a},\tilde{Q}_{\alpha }^{+}\right] &=&-\frac{1}{2}\left(
\gamma _{a}\right) _{\alpha }^{\text{ }\beta }\tilde{Q}_{\beta
}^{-}\,, \quad \ 
\left[ \tilde{G}_{a},\tilde{Q}_{\alpha }^{-}\right] =-\frac{%
1}{2}\left( \gamma _{a}\right) _{\alpha }^{\text{ }\beta }\tilde{R}_{\beta
}\,, \quad \ \ \, \left[ \tilde{P}_{a},\tilde{Q}_{\alpha }^{+}\right] =-\frac{%
1}{2\ell}\left( \gamma _{a}\right) _{\alpha }^{\text{ }\beta }\tilde{Q}_{\beta
}^{-}\,, \notag \\
\left[ \tilde{P}_{a},\tilde{Q}_{\alpha }^{-}\right] &=&-\frac{%
1}{2\ell}\left( \gamma _{a}\right) _{\alpha }^{\text{ }\beta }\tilde{R}_{\beta
}\,, \quad \, 
\left[ \tilde{Z}_{a},\tilde{Q}_{\alpha }^{+}\right] =-\frac{%
1}{2\ell^{2}}\left( \gamma _{a}\right) _{\alpha }^{\text{ }\beta }\tilde{Q}_{\beta
}^{-}\,, \ \ \, \left[ \tilde{Z}_{a},\tilde{Q}_{\alpha }^{-}\right] =-\frac{%
1}{2\ell^{2}}\left( \gamma _{a}\right) _{\alpha }^{\text{ }\beta }\tilde{R}_{\beta
}\,,  \notag \\
\left[ \tilde{Y}_{1},\tilde{Q}_{\alpha }^{+}\right] &=&\frac{1}{2}\left(
\gamma _{0}\right) _{\alpha \beta }\tilde{Q}_{\beta }^{+}\,,\qquad \, \left[ \tilde{Y}_{1},\tilde{Q}_{\alpha }^{-}\right] =-\frac{1}{2}\left(
\gamma _{0}\right) _{\alpha \beta }\tilde{Q}_{\beta }^{-}\,, \quad \ \  \left[ \tilde{Y}_{1},\tilde{R}_{\alpha }\right] =\frac{1}{2}\left(
\gamma _{0}\right) _{\alpha \beta }\tilde{R}_{\beta }\,,  \notag\\
\left[ \tilde{Y}_{2},\tilde{Q}_{\alpha }^{+}\right] &=&\frac{1}{2}\left(
\gamma _{0}\right) _{\alpha \beta }\tilde{R}_{\beta }\,, \qquad \, \left[ \tilde{U}_{1},\tilde{Q}_{\alpha }^{+}\right] =\frac{1}{2\ell^{2}}\left(
\gamma _{0}\right) _{\alpha \beta }\tilde{Q}_{\beta }^{+}\,, \quad \,  \, \left[ \tilde{U}_{1},\tilde{Q}_{\alpha }^{-}\right] =-\frac{1}{2\ell^{2}}\left(
\gamma _{0}\right) _{\alpha \beta }\tilde{Q}_{\beta }^{-}\,,  \notag\\
\left[ \tilde{U}_{1},\tilde{R}_{\alpha }\right] &=&\frac{1}{2\ell^{2}}\left(
\gamma _{0}\right) _{\alpha \beta }\tilde{R}_{\beta }\,, \quad \  \left[ \tilde{U}_{2},\tilde{Q}_{\alpha }^{+}\right] =\frac{1}{2\ell^{2}}\left(
\gamma _{0}\right) _{\alpha \beta }\tilde{R}_{\beta }\,,\label{sEEB2a}
\end{eqnarray}
while the fermionic generators satisfy the following anti-commutation relations:
\begin{eqnarray}
\left\{ \tilde{Q}_{\alpha }^{+},\tilde{Q}_{\beta }^{+}\right\} &=&-\frac{1}{\ell}\left(
\gamma ^{0}C\right) _{\alpha \beta }\tilde{H}-\left(
\gamma ^{0}C\right) _{\alpha \beta }\tilde{Z}-\left( \gamma ^{0}C\right)
_{\alpha \beta }\left(\tilde{U}_{1}+\frac{1}{\ell}\tilde{B}_{1}\right)\,,\notag \\
\left\{ \tilde{Q}_{\alpha }^{+},\tilde{Q}_{\beta }^{-}\right\} &=&-\frac{1}{\ell}\left(
\gamma ^{a}C\right) _{\alpha \beta }\tilde{P}_{a}-\left(
\gamma ^{a}C\right) _{\alpha \beta }\tilde{Z}_{a}\,, \notag\\
\left\{ \tilde{Q}_{\alpha }^{+},\tilde{R}_{\beta }\right\} &=&-\frac{1}{\ell}\left(
\gamma ^{0}C\right) _{\alpha \beta }\tilde{M}-\left(
\gamma ^{0}C\right) _{\alpha \beta }\tilde{T}-\left( \gamma ^{0}C\right)
_{\alpha \beta }\left(\tilde{U}_{2}+\frac{1}{\ell}\tilde{B}_{2}\right)\,, \notag\\
\left\{ \tilde{Q}_{\alpha }^{-},\tilde{Q}_{\beta }^{-}\right\} &=&-\frac{1}{\ell}\left(
\gamma ^{0}C\right) _{\alpha \beta }\tilde{M}-\left(
\gamma ^{0}C\right) _{\alpha \beta }\tilde{T}+\left( \gamma ^{0}C\right)
_{\alpha \beta }\left(\tilde{U}_{2}+\frac{1}{\ell}\tilde{B}_{2}\right)\,.\label{sEEB2b}
\end{eqnarray}
The expanded NR superalgebra corresponds to an alternative non-standard supersymmetric extension of the EEB algebra introduced in \cite{Concha:2019lhn}. Unlike the previous EEB superalgebra obtained in this work, the NR centrally extended superalgebra \eqref{sEEB2a}-\eqref{sEEB2b} does not contain additional fermionic generators, reducing considerably the number of (anti-)commutators. One can notice that $B_1$ and $B_2$ are central charges, while $\{Y_1,Y_2\}$ and $\{U_1,U_2\}$ correspond to expansions of the relativistic R-symmetry generators $\mathcal{T}$ and $\mathcal{U}$, respectively. 
Interestingly, both supersymmetric descriptions of the EEB algebra can be contracted through a vanishing cosmological constant limit $\ell\rightarrow\infty$. As we have shown, the flat limit applied to the EEB superalgebra given by \eqref{EEB1}, \eqref{SEEBa}, \eqref{SEEBb}, and \eqref{SEEBc} reproduces the MEB superalgebra \cite{Concha:2019mxx} allowing to construct a consistent NR supergravity which generalizes the extended Bargmann supergravity \cite{Bergshoeff:2016lwr}. On the other hand, although a flat limit can be done at the level of the non-standard EEB superalgebra \eqref{sEEB2a}-\eqref{sEEB2b}, the resulting NR superalgebra does not guarantee the proper construction of a NR supergravity action. Similarly to the non-standard relativistic Maxwell superalgebra, the absence of the $P_a$ generators in the anti-commutator after considering $\ell\rightarrow\infty$ would reproduce an exotic NR supersymmetric CS action. It would be interesting to avoid such difficulty by exploring a NR version of the non-standard Maxwell algebra using the Hietarinta basis \cite{Hietarinta:1975fu}.

One can notice that the EEB superalgebra \eqref{sEEB2a}-\eqref{sEEB2b} can be written alternatively as the direct sum of three copies of the Nappi-Witten algebra \cite{Nappi:1993ie,Figueroa-OFarrill:1999cmq} where one copy is supersymmetric. Indeed, one can consider the following redefinition of the generators,
\begin{eqnarray}
\tilde{G}_a&=&\hat{G}_a+{G}_{a}+{G}_{a}^{\ast}\,,\qquad \tilde{P}_{a}=\frac{1}{\ell}\left(G_{a}-G_{a}^{\ast}\right)\,, \qquad \tilde{Z}_{a}=\frac{1}{\ell^{2}}\left(G_{a}+G_{a}^{\ast} \right)\,, \notag\\
\tilde{S}&=&\hat{S}+{S}+{S}^{\ast}\,,\qquad \ \ \ \ \tilde{M}=\frac{1}{\ell}\left(S-S^{\ast}\right)\,, \qquad \ \ \ \, \tilde{T}=\frac{1}{\ell^{2}}\left(S+S^{\ast} \right)\,, \notag\\
\tilde{J}&=&\hat{J}+{J}+{J}^{\ast}\,,\qquad \ \ \ \ \, \tilde{H}=\frac{1}{\ell}\left(J-J^{\ast}\right)\,, \qquad \ \ \ \,  \tilde{Z}=\frac{1}{\ell^{2}}\left(J+J^{\ast} \right)\,, \notag\\
\tilde{Y}_1&=&\hat{T}_{1}+{T}_{1}+{T}_{1}^{\ast}\,,\qquad  \ \, \tilde{B}_1=\frac{1}{\ell}\left(T_{1}-T_{1}^{\ast}\right)\,, \qquad \   \tilde{U}_1=\frac{1}{\ell^{2}}\left(T_{1}+T_{1}^{\ast} \right)\,, \notag\\
\tilde{Y}_2&=&\hat{T}_{2}+{T}_{2}+{T}_{2}^{\ast}\,,\qquad \  \, \tilde{B}_2=\frac{1}{\ell}\left(T_{2}-T_{2}^{\ast}\right)\,, \qquad \   \tilde{U}_2=\frac{1}{\ell^{2}}\left(T_{2}+T_{2}^{\ast} \right)\,, \notag\\
\tilde{Q}_{\alpha}^{+}&=&\frac{\sqrt{2}}{\ell}\mathcal{Q}_{\alpha}^{+}\,,\qquad \qquad \quad   \tilde{Q}_{\alpha}^{-}=\frac{\sqrt{2}}{\ell}\mathcal{Q}_{\alpha}^{-}\,, \qquad \qquad  \,  \tilde{R}_{\alpha}=\frac{\sqrt{2}}{\ell}\mathcal{R}_{\alpha}\,, \label{redef22}
\end{eqnarray}
and see that both subsets spanned by $\{\hat{G}_{a},\hat{S},\hat{J} \}$ and $\{G_{a}^{\ast},S^{\ast},J^{\ast} \}$ define a Nappi-Witten algebra \eqref{NW} coupled to $\mathfrak{u}(1)$ generators $\{\hat{T}_{1},\hat{T}_{2}\}$ and $\{T_{1}^{\ast},T_{2}^{\ast} \}$, respectively. On the other hand, the set of generators $\{ G_a,S,J,T_{1},T_{2},\mathcal{Q}_{\alpha}^{+},\mathcal{Q}_{\alpha}^{-},\mathcal{R}_{\alpha} \}$ satisfy a supersymmetric extension of the Nappi-Witten algebra whose (anti-)commutation relations are given by \eqref{sNW}. It is interesting to point out that, although the non-standard EEB superalgebra seems to be quite different to the ``standard" one studied previously, they actually differ just on the amount of supersymmetric copies of the Nappi-Witten algebra.

An alternative change of basis allows us to rewrite the EEB superlalgebra \eqref{sEEB2a}-\eqref{sEEB2b} as the direct sum of the extended Newton-Hooke superalgebra and the Nappi-Witten algebra. Such structure can be obtained by considering the following redefinition of the generators:
\begin{eqnarray}
\tilde{G}_{a}&=&G_{a}^{\star}+G_{a}\,, \qquad \qquad \tilde{P}_{a}=P_{a}\,,\qquad \qquad \tilde{Z}_{a}=\frac{1}{\ell^{2}}G_{a}\,, \notag \\
\tilde{S}&=&S^{\star}+S\,, \qquad \qquad \ \  \tilde{M}=M\,,\qquad \qquad \, \  \tilde{T}=\frac{1}{\ell^{2}}S\,, \notag \\
\tilde{J}&=&J^{\star}+J\,, \qquad \qquad \ \ \,  \tilde{H}=H\,,\qquad \qquad \ \  \tilde{Z}=\frac{1}{\ell^{2}}J\,, \notag \\
\tilde{Y}_{1}&=&T_{1}^{\star}+T_{1}\,, \qquad \qquad  \,  \tilde{B}_{1}=B_1\,,\qquad \qquad  \  \tilde{U}_{1}=\frac{1}{\ell^{2}}T_{1}\,, \notag \\
\tilde{Y}_{2}&=&T_{2}^{\star}+T_{2}\,, \qquad \qquad \,   \tilde{B}_{2}=B_2\,,\qquad \qquad  \  \tilde{U}_{2}=\frac{1}{\ell^{2}}T_{2}\,, \notag \\
\tilde{Q}_{\alpha}^{+}&=& \sqrt{\frac{1}{\ell}}\mathcal{Q}_{\alpha}^{+}\,,\qquad \qquad \  \tilde{Q}_{\alpha}^{-}=\sqrt{\frac{1}{\ell}}\mathcal{Q}_{\alpha}^{-}\,, \quad \quad \  \tilde{R}_{\alpha}=\sqrt{\frac{1}{\ell}}\mathcal{R}_{\alpha}\,. \label{redef3}
\end{eqnarray}

The subset spanned by $\{G_{a},P_{a},S,M,J,H,T_1,T_2,\mathcal{Q}_{\alpha}^{+},\mathcal{Q}_{\alpha}^{-},\mathcal{R}_{\alpha} \}$ corresponds to the extended Newton-Hooke superalgebra \eqref{sENH} (with, in this case, $U_1=U_2=0$) now endowed with a central extension given by $B_1$ and $B_2$. On the other hand, the set of generators $\{G_{a}^{\star},S^{\star},J^{\star},T_{1}^{\star},T_{2}^{\star} \}$ satisfies the Nappi-Witten algebra \eqref{NW} coupled to the $\mathfrak{u}(1)$ generators $T_{1}^{\star}$ and $T_{2}^{\star}$. Here, we shall focus on the EEB superalgebra written as in \eqref{sEEB2a}-\eqref{sEEB2b} which, as we shall see, offers us an alternative way to introduce a cosmological constant in NR supergravity theory different from the one presented in the previous section and from the extended Newton-Hooke one \cite{Ozdemir:2019tby}.

\subsection{Non-standard non-relativistic extended supergravity action}

We can now construct a NR CS supergravity action based on the non-standard EEB superalgebra previously introduced. To this aim, let us start by writing the non-vanishing components of the invariant tensor
for the non-standard EEB superalgebra, which can be determined by exploiting the $S$-expansion on \eqref{nsinvt} (see Appendix \ref{appa}). In particular, the non-standard EEB superalgebra admits the bosonic invariant tensor \eqref{invt1} along with the following non-vanishing components of the invariant tensor
\begin{eqnarray}
\left\langle \tilde{Y}_{1}\tilde{Y}_{2}\right\rangle &=&\tilde{\alpha}_{0}\,, \qquad \qquad \ \ \left\langle \tilde{Y}_{1}\tilde{B}_{2}\right\rangle =\tilde{\alpha}_{1}\,, \qquad \qquad \ \  \left\langle \tilde{Y}_{2}\tilde{B}_{1}\right\rangle = \tilde{\alpha}_{1} \,,  \notag \\
\left\langle \tilde{Y}_{1}\tilde{U}_{2}\right\rangle &=&\tilde{\alpha}_{2}\,, \qquad \qquad \ \ \left\langle \tilde{Y}_{2}\tilde{U}_{1}\right\rangle = \tilde{\alpha}_{2}\,, \qquad \qquad \ \ \left\langle \tilde{U}_{1}\tilde{B}_{2}\right\rangle = \frac{\tilde{\alpha}_{1}}{\ell^{2}}\,,\notag \\
\left\langle \tilde{U}_{2}\tilde{B}_{1}\right\rangle &=&\frac{\tilde{\alpha}%
_{1}}{\ell ^{2}}\,, \qquad \qquad \ \left\langle \tilde{B}_{1}\tilde{B}_{2}\right\rangle = \frac{\tilde{\alpha}_{1}}{\ell^{2}}\,, \qquad \qquad \ \left\langle \tilde{U}_{1}\tilde{U}_{2}\right\rangle =\frac{\tilde{\alpha}%
_{2}}{\ell ^{2}}\,,  \notag \\
\left\langle \tilde{Q}_{\alpha }^{-}\tilde{Q}_{\beta }^{-}\right\rangle &=&2%
\left(\frac{\tilde{\alpha}_{1}}{\ell}+\tilde{\alpha}_{2} \right)C_{\alpha \beta }\,, \notag \\ \left\langle \tilde{Q}_{\alpha }^{+}%
\tilde{R}_{\beta }\right\rangle &=& 2%
\left(\frac{\tilde{\alpha}_{1}}{\ell}+\tilde{\alpha}_{2} \right)C_{\alpha \beta }\,, \label{invtsEEB}
\end{eqnarray}
where $\tilde{\alpha}_0$, $\tilde{\alpha}_1$ and $\tilde{\alpha}_2$ are arbitrary constants which are related to the relativistic ones through \eqref{sexp2a}.

The gauge connection one-form $A$ for the non-standard EEB superalgebra is
\begin{eqnarray}
A &=&\omega \tilde{J}+\omega ^{a}\tilde{G}_{a}+\tau \tilde{H}+e^a
\tilde{P}_{a}+k\tilde{Z}+k^{a}\tilde{Z}_{a}+m\tilde{M}+s\tilde{S}+t\tilde{T}+y_{1}\tilde{Y_{1}}+y_{2}\tilde{Y}_{2}
\notag \\
&&+b_1\tilde{B}_{1}+b_{2}\tilde{B}%
_{2}+u_{1}\tilde{U}_{1}+u_{2}\tilde{U}_{2}+\bar{\psi }^{+}\tilde{Q}^{+}+\bar{\psi }^{-}\tilde{Q}^{-}+\bar{\rho }\tilde{R} \,,
\label{oneformNSEEB}
\end{eqnarray}
and the corresponding curvature two-form is given by
\begin{eqnarray}
F &=&F\left( \omega \right) \tilde{J}+F^{a}\left( \omega ^{b}\right)
\tilde{G}_{a}+F\left( \tau \right) \tilde{H}+F^{a}\left( e^{b}\right) \tilde{%
P}_{a}+F\left( k\right) \tilde{Z}+F^{a}\left( k^{b}\right) \tilde{Z}%
_{a}+F\left( m\right) \tilde{M}  \notag \\
&&+F\left( s\right) \tilde{S}+F\left( t\right) \tilde{T}+F\left(
y_{1}\right) \tilde{Y_{1}}+F\left( y_{2}\right) \tilde{Y}_{2}+F\left(
b_{1}\right) \tilde{B}_{1}+F\left( b_{2}\right) \tilde{B}_{2}+F\left(
u_{1}\right) \tilde{U}_{1}  \notag \\
&&+F\left( u_{2}\right) \tilde{U}_{2}+\nabla  \bar{\psi }^{+}\tilde{Q}^{+}+\nabla  \bar{\psi }^{-}\tilde{Q}^{-}+\nabla  \bar{\rho}\tilde{R} \,, \label{F2cNSEEB}
\end{eqnarray}
where
\begin{eqnarray}
F\left( \omega \right) &=&R\left( \omega \right) \,,  \notag \\
F^{a}\left( \omega ^{b}\right) &=&R^{a}\left( \omega ^{b}\right) \,,  \notag
\\
F\left( \tau \right) &=&R\left( \tau \right) +\frac{1}{2 \ell}\bar{\psi}%
^{+}\gamma ^{0}\psi ^{+} \,,  \notag \\
F^{a}\left( e^{b}\right) &=&R^{a}\left( e^{b}\right) +\frac{1}{\ell}\bar{\psi}^{+}\gamma
^{a}\psi ^{-} \,,  \notag
\\
F\left( k\right) &=&R\left( k\right) +\frac{1}{2}\bar{\psi}^{+}\gamma ^{0}\psi ^{+}\,,
\notag \\
F^{a}\left( k^{b}\right) &=&R^{a}\left( k^{b}\right) +\bar{\psi}^{+}\gamma
^{a}\psi ^{-} \,,  \notag \\
F\left( m\right) &=&R\left( m\right) +\frac{1}{\ell}\bar{\psi}^{+}\gamma ^{0}\rho + \frac{1}{2 \ell}\bar{\psi}^{-}\gamma^{0}\psi ^{-} \,,
\notag \\
F\left( s\right) &=&R\left( s\right) \,,  \notag \\
F\left( t\right) &=&R\left( t\right) +\bar{\psi}^{+}\gamma ^{0}\rho +\frac{1}{2} \bar{\psi}^{-}\gamma ^{0}\psi ^{-} \,,  \label{boscurvNSEEBp1}
\end{eqnarray}
being $R\left( \omega \right) $, $R^{a}\left( \omega ^{b}\right) $, $R\left(
\tau \right) $, $R^{a}\left( e^{b}\right) $, $R\left( k\right) $, $%
R^{a}\left( k^{b}\right) $, $R\left( m\right) $, $R\left( s\right) $, and $%
R\left( t\right) $ the bosonic EEB curvatures defined in (\ref{curvEEB}),
together with
\begin{eqnarray}
F\left( y_{1}\right) &=&dy_{1} \,,  \notag \\
\hat{F}\left( y_{2}\right) &=&dy_{2} \,,  \notag \\
F\left( b_{1}\right) &=&db_{1}+ \frac{1}{2\ell}\bar{\psi}^{+}\gamma ^{0}\psi ^{+}\,,  \notag
\\
F\left( b_{2}\right) &=&db_{2}+\frac{1}{\ell}\bar{\psi}%
^{+}\gamma ^{0}\rho - \frac{1}{2\ell}\bar{\psi}^{-}\gamma ^{0}\psi^- \,,
\notag \\
F\left( u_{1}\right) &=&du_{1}+\frac{1}{2}\bar{\psi}^{+}\gamma ^{0}\psi ^{+} \,,  \notag \\
F\left( u_{2}\right) &=&du_{2} + \bar{\psi}^+ \gamma^0 \rho -\frac{1}{2}\bar{\psi}^{-}\gamma ^{0}\psi ^{-} \,.  \label{boscurvNSEEBp2}
\end{eqnarray}
On the other hand, the covariant derivatives of the spinor $1$-form fields read
\begin{eqnarray}
\nabla \psi ^{+} &=&d\psi ^{+}+\frac{1}{2}\omega \gamma _{0}\psi ^{+}-\frac{1%
}{2}y_{1}\gamma _{0}\psi ^{+}+\frac{1}{2\ell}\tau \gamma _{0}\psi ^{+}+%
\frac{1}{2\ell ^{2}}k\gamma _{0}\psi ^{+}-\frac{1}{2\ell ^{2}}u_{1}\gamma
_{0}\psi ^{+} \,,  \notag \\
\nabla  \psi ^{-} &=&d\psi ^{-}+\frac{1}{2}\omega \gamma _{0}\psi ^{-}+\frac{1%
}{2}\omega ^{a}\gamma _{a}\psi ^{+}+\frac{1}{2}y_{1}\gamma _{0}\psi ^{-}+%
\frac{1}{2\ell}\tau \gamma _{0}\psi ^{-}+\frac{1}{2\ell}e^{a}\gamma
_{a}\psi ^{+}+\frac{1}{2\ell ^{2}}k^{a}\gamma _{a}\psi ^{+}  \notag \\
&&+\frac{1}{2\ell ^{2}}k\gamma _{0}\psi ^{-}+\frac{1}{2\ell ^{2}}u_{1}\gamma
_{0}\psi ^{-} \,, \notag \\
\nabla  \rho &=&d\rho +\frac{1}{2}\omega \gamma _{0}\rho +\frac{1}{2}\omega
^{a}\gamma _{a}\psi ^{-}+\frac{1}{2}s\gamma _{0}\psi ^{+}-\frac{1}{2}%
y_{2}\gamma _{0}\psi ^{+}-\frac{1}{2}y_{1}\gamma _{0}\rho +\frac{1}{2\ell}e^{a}\gamma _{a}\psi ^{-}  \notag \\
&&+\frac{1}{2\ell ^{2}}k^{a}\gamma _{a}\psi ^{-}+\frac{1}{2\ell}m\gamma
_{0}\psi ^{+}+\frac{1}{2\ell}\tau \gamma _{0}\rho +\frac{1}{2\ell ^{2}}%
t\gamma _{0}\psi ^{+}+\frac{1}{2\ell ^{2}}k\gamma _{0}\rho  \notag \\
&&-\frac{1}{2\ell ^{2}}u_{1}\gamma _{0}\rho -\frac{1}{2\ell ^{2}}u_{2}\gamma
_{0}\psi ^{+} \,.  \label{fermcurvNSEEB}
\end{eqnarray}

A CS supergravity action based on the non-standard (NS) EEB superalgebra \eqref{sEEB2a}-\eqref{sEEB2b} can be constructed by combining the non-zero components of the
invariant tensor \eqref{invt1} and \eqref{invtsEEB} with
the gauge connection one-form $A$ (\ref{oneformNSEEB}). One can see that the explicit CS supergravity action can be split in three independent sectors:
\begin{equation}
I_{\text{NR}}^{\text{NS}}=\tilde{\alpha}_{0}\tilde{I}_{0}+\tilde{\alpha}_{1}\tilde{I}_{1}+\tilde{\alpha}_{2}\tilde{I}_{2}\,,
\end{equation}
where
\begin{eqnarray}
\tilde{I}_{0}&=&\int \omega_a R^{a}\left(\omega ^{b}\right)-2sR\left(\omega \right) +2y_{1}dy_{2}\,, \notag \\
\tilde{I}_{1}&=&\int  e_a R^a\left(\omega ^{b}\right) + \omega_a R^a\left(e^b\right) - 2 m R(\omega) - 2 \tau ds -\frac{2}{\ell^2} m R(k) - \frac{2}{\ell^2} \tau dt + \frac{1}{\ell^2} e_a R^a\left(k^b\right) \notag \\
&& + \frac{1}{\ell^2} k_a R^a \left(e^b\right)+ 2 b_1 dy_1 + 2 b_1 dy_2 + \frac{2}{\ell^2} u_1 db_2 + \frac{2}{\ell^2} u_2 db_1 - \frac{2}{\ell} b_1 db_2 - \frac{2}{\ell} \bar{\psi}^- \nabla  \psi^-  \notag \\
&&- \frac{2}{\ell} \bar{\psi}^+ \nabla  \rho-\frac{2}{\ell} \bar{\rho} \nabla  \psi^+\,,
\notag \\
\tilde{I}_{2}&=&\int \omega_a R^a\left(k^b\right) + k_a R^a\left(\omega^b\right)+ e_a R^a\left(e^b\right) - 2 s R(k) - 2 t R(\omega) - 2 m R(\tau)+ \frac{1}{\ell^2} k_a R\left(k^b\right)  \notag \\
&&   - \frac{2}{\ell ^{2}} t R(k) + 2 u_2 dy_1 + 2 u_1 dy_2+ \frac{2}{\ell^2} u_1 du_2 - 2 \bar{\psi}^- \hat{\nabla}  \psi^- - 2 \bar{\psi}^+ \hat{\nabla}  \rho - 2 \bar{\rho} \hat{\nabla}  \psi^+ \,. \label{CS3}
\end{eqnarray}
The CS action (\ref{CS3}) obtained here describes a non-standard Enlarged
Extended Bargmann supergravity model whose field equations are given by the vanishing of the curvature 2-form \eqref{F2cNSEEB}. Although it contains the bosonic EEB gravity theory, the non-standard version of the EEB superalgebra leads to a CS action diverse from the ``standard" one \eqref{CS2} discussed in the previous section. Both supersymmetric descriptions contain the exotic NR gravity action $\tilde{I}_{0}=I_{0}$ as particular subcase. However, $\tilde{I}_{1}$ and $\tilde{I}_{2}$ are quite different from the $I_{1}$ and $I_{2}$ contributions appearing in the standard case. Indeed, the fact that the non-standard EEB superalgebra contains only three different spinor generators reduces considerably the CS expression. Furthermore, the vanishing cosmological constant limit $\ell\rightarrow\infty$ in the non-standard case reproduces an exotic supersymmetric theory. In particular, the flat limit applied to the $\tilde{I}_{1}$ contribution does not reproduce a supergravity theory anymore, but leads us to the extended Bargmann gravity \cite{Bergshoeff:2016lwr} coupled to extra generators $\{y_1,y_2,b_1,b_2\}$. Such peculiar behavior is inherited from the relativistic counterpart and is responsible of the label ``non-standard". However, a different interpretation of the gauge field could overcome such difficulty to reproduce a supergravity theory in the flat limit by using the Hietarinta-Maxwell interpretation \cite{Chernyavsky:2020fqs}. It would be interesting to explore the Physical implications of considering the Hietarinta basis at both the relativistic and the NR levels.

On the other hand, as in the ``standard" case, the CS supergravity action presented here is restricted exclusively to the CS terms related to the $\lambda_2$ element of the semigroup $S_{E}^{(2)}$. The exotic-like terms produced due to the $\lambda_0$ element has been omitted intentionally. Indeed, such terms would modify the CS bosonic sector, meaning that one would have a supersymmetric extension of a diverse CS gravity theory.

An alternative procedure to recover the non-standard EEB supergravity action \eqref{CS3} can be performed by considering an $S^{(2)}_E$-expansion of the relativistic $\mathcal{N}=2$ non-standard AdS-$\mathcal{L}$ CS supergavity action,\footnote{The relativistic CS action appears considering the non-vanishing components of the invariant tensor for the $\mathcal{N}=2$ AdS-$\mathcal{L}$ superalgebra \eqref{nsinvt}.}
\begin{eqnarray}
I_{\text{R}}^{\text{NS}}&=&\int \alpha_0\left[\omega_A d\omega^{A}+\frac{1}{3}\epsilon_{ABC}\omega^{A}\omega^{B}\omega^{C} +\mathsf{t}d\mathsf{t} \right] +\alpha_{1}\left[2e_{A}R^{A}+\frac{1}{3\ell^{2}}\epsilon_{ABC}e^{A}e^{B}e^{C}+\frac{2}{\ell^{2}}e_{A}F^{A} \right. \notag \\
&&\left.-\frac{2}{\ell}\bar{\hat{\psi}}^{i}\nabla\hat{\psi}^{i}+\mathsf{t}d\mathsf{b}+\frac{1}{\ell^{2}}\mathsf{u}d\mathsf{b}+\frac{1}{\ell}\mathsf{b}d\mathsf{b} \right]+\alpha_2 \left[ 2\sigma_AR^{A}+e_AT^{A}+\frac{2}{\ell^{2}}e_AF^{A}+\frac{1}{\ell^{2}}\epsilon_{ABC}e^{A}\sigma^{B}e^{C}\right.\notag \\
&&\left. -2\bar{\hat{\psi}}^{i}\nabla\hat{\psi}^{i}+\mathsf{t}d\mathsf{u}+\frac{1}{\ell^{2}}\mathsf{u}d\mathsf{u}\right]\,,\label{N2NSADSL}
\end{eqnarray}
where $R^{A}$, $T^{A}$, and $F^{A}$ are defined as in \eqref{relcurv}, while the covariant derivative of the fermionic gauge field is given by
\begin{equation}
\nabla\hat{\psi}^{i}=d\hat{\psi}^{i}+\frac{1}{2}\omega^{A}\gamma_{A}\hat{\psi}^{i}+\frac{1}{2\ell}e^{A}\gamma_{A}\hat{\psi}^{i}+\frac{1}{2\ell^{2}}\sigma^{A}\gamma_{A}\hat{\psi}+\mathsf{t}\epsilon^{ij}\hat{\psi}^{j}+\frac{1}{\ell^{2}}\mathsf{u}\epsilon^{ij}\hat{\psi}^{j}\,. \label{relcurv2}
\end{equation}
In fact, the NR gauge fields can be written in terms of the relativistic ones and the semigroup elements as
\begin{eqnarray}
\omega&=&\lambda_0 \omega_0\,, \qquad  \ \, s=\lambda_2 \omega_0\,, \qquad \omega_a=\lambda_1 \omega_a\,, \notag \\
\tau&=&\lambda_0 e_0\,, \qquad \ m=\lambda_2 e_0\,, \qquad \ e_a=\lambda_1 e_a \,, \notag \\
k&=&\lambda_0 \sigma_0\,, \qquad \ \ \, t=\lambda_2 \sigma_0\,, \qquad  \, k_a=\lambda_1 \sigma_a\,, \notag \\
\psi_{\alpha}^{+}&=&\lambda_0 \hat{\psi}_{\alpha}^{+}\,, \quad \ \  \rho_{\alpha}=\lambda_2 \hat{\psi}_{\alpha}^{+}\,, \quad \ \, \psi_{\alpha}^{-}=\lambda_1 \hat{\psi}_{\alpha}^{-}\,, \notag \\
y_{1}&=&\lambda_0 \mathsf{t}\,, \qquad \ \ u_1=\lambda_0 \mathsf{u}\,, \qquad \ \   b_1=\lambda_0 \mathsf{b}\,, \notag\\
y_{2}&=&\lambda_2 \mathsf{t}\,, \qquad \ \  u_2=\lambda_2 \mathsf{u}\,, \qquad \ \  b_2=\lambda_2 \mathsf{b}\,,\label{gfredef3}
\end{eqnarray}
where we have performed the split $A=\lbrace 0,a \rbrace$ with $a=1,2$ and where $\hat{\psi}^{\pm}_{\alpha}$ is defined as in \eqref{redeff2}. Thus, the non-standard EEB supergravity action can be obtained by considering the expanded gauge fields \eqref{gfredef3}, the expanded parameters \eqref{sexp2a}, the multiplication law \eqref{ml}, together with the $0_s$-reduction property. 


\section{Discussion}

In this work we have presented two diverse supersymmetric extensions of the enlarged extended Bargmann gravity in three spacetime dimensions introduced in \cite{Concha:2019lhn}. The new NR superalgebras we have introduced differ on the number of fermionic generators and have been obtained from two different relativistic $\mathcal{N}=2$ AdS-$\mathcal{L}$ superalgebras considering the semigroup expansion procedure \cite{Izaurieta:2006zz}. We have shown that both EEB superalgebras allows us to introduce a cosmological constant to a NR supergravity different from the extended Newton-Hooke one\cite{Ozdemir:2019tby}. Nevertheless, only the ``standard" description reproduces a consistent vanishing cosmological constant limit $\ell\rightarrow\infty$ leading to the Maxwellian version of the extended Bargmann supergravity presented in \cite{Concha:2019mxx}. Indeed, the flat limit of the non-standard version of the EEB supergravity leads us to an exotic supersymmetric theory since $\{ Q,Q\}\sim Z_{a}$. One way to overcome such difficulty could be performed by interchanging the role of the $P_{a}$ and $Z_{a}$ generators as in the relativistic Hietarinta-Maxwell gravity theory \cite{Chernyavsky:2020fqs}. Motivated by the fact that both topological and minimal massive gravity theories \cite{Deser:1981wh, Bergshoeff:2014pca} appear as particular cases of a generalized minimal massive gravity arising from a spontaneous breaking of the Hietarinta-Maxwell in a CS theory, it would be worth it to explore the Physical implications of a possible NR version of the three-dimensional Hietarianta-Maxwell CS (super)gravity.

On the other hand, both EEB superalgebras can be written in terms of three copies of the Nappi-Witten algebra where one or two copies are augmented by supersymmetry depending on the case. An alternative redefinition of the generators reveals that the (non-)standard EEB superalgebra can also be written as the direct sum of the extended Newton-Hooke superalgebra\cite{Ozdemir:2019tby} and the Nappi-Witten (super)algebra.  Such structures are inherited from their relativistic counterparts, in which case the AdS-$\mathcal{L}$ superalgebra are isomorphic to three copies of the Lorentz (super)algebra and to the the direct sum of the AdS superalgebra and the (super) Lorentz one. It would be interesting to explore if the supersymmetric extension of the Nappi-Witten algebra obtained here can be used to obtain novel superalgebras through the $S$-expansion procedure \cite{Concha:2020NNN}. In particular, one could extend the results obtained in \cite{Concha:2019lhn,Penafiel:2019czp,Concha:2020sjt} to the supersymmetric case.

The $S$-expansion method used here is a powerful tool to obtain new NR superalgebras and to construct the respective NR CS supergravity theories. Consequently, the procedure performed here could serve as a starting point for diverse further studies. In particular, we have shown that the $S$-expansion of a relativistic $\mathcal{N}=2$ AdS-$\mathcal{L}$ superalgebra using $S_{E}^{(2)}$ as the relevant semigroup reproduces its NR counterpart. It would be interesting to first check if known NR superalgebras appears from $\mathcal{N}=2$ superalgebras considering the same approach. Then, one could go further and study which kind of superalgebras can be obtained by considering $S_{E}
^{(N)}$-expansions for $N>2$ to diverse relativistic superalgebras. On the other hand, one could apply our method to four and higher spacetime dimensions and construct NR supergravity models. 

Another aspect that deserves further developments is the study of the asymptotic symmetry of the EEB (super)gravity theory. At the relativistic level, a semi-simple enlargement of the $\mathfrak{bms}_{3}$ algebra results to describe the boundary dynamics of the three-dimensional AdS-$\mathcal{L}$ gravity \cite{Concha:2018jjj}. Interestingly, such asymptotic structure can be written as three copies of the Virasoro algebra or as the direct sum of the conformal symmetry and the Virasoro one \cite{Caroca:2017onr,Concha:2019eip}. A future development could consist on the study of consistent boundary conditions for the EEB gauge field to unveil the asymptotic symmetry of the theory. One could expect to obtain a canonical realization of a new infinite-dimensional symmetry which could be written as three copies of the NR version of the Virasoro algebra. 

Besides, since these NR supersymmetry algebras share the presence of extra fermionic generators, it could be intriguing to study higher-dimensional cases and, in particular, to carry on an analysis on their possible hidden gauge structure in higher dimensions, on the same lines of \cite{Andrianopoli:2016osu,Andrianopoli:2017itj,Ravera:2018vra}. This could give new insights in the NR regime of supergravity theories. In correspondence with the present results, it would be worth to analyze the NR counterpart of \cite{Ravera:2018vra} in the presence of a cosmological constant. 

Concerning other possible future developments, it would also be interesting to extend the study to the ultra-relativistic regime, where some results in the context of supergravity have been recently presented in \cite{Ravera:2019ize, Ali:2019jjp} [work in progress].


\section*{Acknowledgments}

This work was funded by the National Agency for Research and Development ANID (ex-CONICYT) - PAI grant No. 77190078 (P.C.) and the FONDECYT Project N$^{\circ }$3170438 (E.R.). P.C. would like to
thank to the Dirección de Investigación and Vice-rectoría de Investigación
of the Universidad Católica de la Santísima Concepción, Chile, for their
constant support. L.R. would like to thank A. Gamba and F. Dolcini for financial support.


\appendix

\section{Invariant tensor and semigroup expansion method}\label{appa}

The semigroup expansion method, introduced in \cite{Izaurieta:2006zz} and latter studied in \cite{Caroca:2011qs,Andrianopoli:2013ooa,Artebani:2016gwh}, consists in combining the elements of a semigroup $S$ with the structure constant of a Lie (super)algebra $\mathfrak{g}$. The new Lie (super)algebra $\mathfrak{G}=S\times \mathfrak{g}$ is said to be an $S$-expanded (super)algebra whose structure constants are related to the structure constant of the orignal Lie (super)algebra $\mathfrak{g}$ as
\begin{equation}
    C_{(A,\alpha)(B,\beta)}^{\qquad \quad(C,\gamma)}=K_{\alpha\beta}^{\quad\gamma}C_{AB}^{\ C}\,,
\end{equation}
where $K_{\alpha\beta}^{\quad\gamma}$ is the 2-selector which encodes the information from the multiplication law of the semigroup $S$ and satisfy
\begin{equation}
    K_{\alpha\beta}^{\quad\gamma}=\left\{
    \begin{array}{c}
    1\text{\qquad when\,}\gamma=\rho(\alpha,\beta) \,, \\
    0\qquad \text{otherwise. \ \ \ \ \ \ \ \  }
    \end{array}
    \right.
\end{equation}

Interestingly, the $S$-expansion procedure allows us to obtain the invariant tensor of the expanded Lie (super)algebra $\mathfrak{G}$. The invariant tensor is a crucial ingredient for the construction of a CS action based on the expanded Lie (super)algebra.

According to \cite{Izaurieta:2006zz}, let us consider $\mathfrak{g}$ being a Lie (super)algebra spanned by $\{T_{A}\}$ and let $S$ be an abelian semigroup. If the original Lie (super)algebra $\mathfrak{g}$ admits $\left\langle T_{A_{1}}\cdots T_{A_{n}}\right\rangle$ as the invariant tensor, then the expanded Lie (super)algebra $\mathfrak{G}=S\times\mathfrak{g}$ admits the following invariant tensor,
\begin{equation}
    \left\langle T_{A_{1},\alpha_{1}}\cdots T_{A_{n},\alpha_{n}}\right\rangle=\alpha_{\gamma}K_{\alpha_{1}\cdots\alpha_{n}}^{\qquad \gamma}\left\langle T_{A_{1}}\cdots T_{A_{n}}\right\rangle \,, \label{invsexp}
\end{equation}
where the $\alpha_{\gamma}$'s are arbitrary constants. Here $K_{\alpha_{1}\cdots\alpha_{n}}$ is the n-selector for the semigroup $S$ which is defined as
\begin{equation}
    K_{\alpha_{1}\cdots\alpha_{n}}^{\qquad\gamma}=\left\{
    \begin{array}{c}
    1\text{\qquad when\,}\gamma=\rho(\alpha_1,\cdots,\alpha_n)\\
    0\qquad \text{otherwise. \ \ \ \ \ \ \ \ \ \ \ \ \ \ \ \ }
    \end{array}
    \right.
\end{equation}

The possibility to find the invariant tensor for an expanded Lie (super)algebra is an additional advantage of the $S$-expansion method. Furthermore, the $S$-expansion procedure allows us to obtain new expanded Lie (super)algebras which cannot be obtained through other Lie (super)algebra expansion methods. Indeed, an $S$-expansion performed by using the particular semigroup $S_{E}^{(N)}$ reproduces the expanded Lie (super)algebras derived through the Maurer-Cartan forms power series expansion \cite{deAzcarraga:2002xi,deAzcarraga:2007et}. A different choice of the semigroup would lead to other expanded Lie (super)algebras which cannot be obtained through the Maurer-Cartan formalism.


\section{Gauge transformations}\label{appb}

The supersymmetry transformation laws of the EEB superalgebra read
\begin{eqnarray}
\delta \omega &=&0\,, \quad \quad \quad \delta \omega ^{a} \, = \, 0\,,
\quad \quad \quad \delta \tau \, = \, - \bar{\epsilon}%
^{+}\gamma ^{0}\psi ^{+} - \frac{1}{\ell^2} \bar{\varphi}^+
\gamma^0 \xi^+ \,,  \notag \\
\delta e^{a} &=& - \bar{\epsilon}^{+}\gamma ^{a}\psi ^{-} %
- \bar{\epsilon} ^{-}\gamma ^{a}\psi ^{+} -
\frac{1}{\ell^2} \bar{\varphi^+} \gamma^a \xi^- - \frac{1}{%
\ell^2} \bar{\varphi}^- \gamma^a \xi^+ \,,  \notag \\
\delta k &=& - \bar{\epsilon}^{+}\gamma ^{0}\xi ^{+} %
- \bar{\varphi}^{+}\gamma^{0}\psi ^{+}\,,  \notag \\
\delta k^{a} &=& - \bar{\epsilon}^{+ }\gamma ^{a}\xi ^{- } -
\bar{\epsilon}^{- }\gamma ^{a}\xi ^{+ } - \bar{\varphi}^{+ }\gamma ^{a}\psi
^{- } - \bar{\varphi}^{- }\gamma ^{a}\psi ^{+ }  \,,  \notag \\
\delta m &=& - \bar{\epsilon}^{-}\gamma ^{0}\psi ^{-} %
- \bar{\epsilon}^{+}\gamma ^{0}\rho - \bar{%
\eta}\gamma ^{0}\psi ^{+} - \frac{1}{\ell^2} \bar{\varphi}%
^- \gamma^0 \xi^- - \frac{1}{\ell^2} \bar{\varphi}^+
\gamma^0 \chi - \frac{1}{\ell^2} \bar{\zeta} \gamma^0 \xi^+
\,,  \notag \\
\delta s &=&0\,,  \notag \\
\delta t &=& - \bar{\epsilon}^{-}\gamma ^{0}\xi ^{-} %
- \bar{\varphi}^{-}\gamma^{0}\psi ^{-} -
\bar{\epsilon}^{+}\gamma ^{0}\chi - \bar{\zeta}%
\gamma^{0}\psi ^{+} - \bar{\varphi}^{+}\gamma ^{0}\rho %
- \bar{\eta}\gamma ^{0}\xi^{+}\,,  \notag \\
\delta y_1 &=& 0 \,,  \notag \\
\delta y_2 &=& 0 \,,  \notag \\
\delta b_1 &=& - \bar{\epsilon}^+ \gamma^0 \xi^+ %
- \bar{\varphi}^+ \gamma^0 \psi^+ \,,  \notag \\
\delta b_2 &=& \bar{\epsilon}^- \gamma^0 \xi^- %
+ \bar{\varphi}^- \gamma^0 \psi^- - \bar{%
\epsilon}^+ \gamma^0 \chi - \bar{\zeta} \gamma^0 \psi^+ %
- \bar{\varphi}^+ \gamma^0 \rho - \bar{\eta%
} \gamma^0 \xi^+ \,,  \notag \\
\delta u_1 &=& - \bar{\epsilon}^+ \gamma^0 \psi^+ %
- \frac{1}{\ell^2} \bar{\varphi}^+ \gamma^0 \xi^+ \,,
\notag \\
\delta u_2 &=& \bar{\epsilon}^- \gamma^0 \psi^- %
- \bar{\epsilon}^+ \gamma^0 \rho - \bar{%
\eta} \gamma^0 \psi^+ + \frac{1}{\ell^2} \bar{\varphi}^-
\gamma^0 \xi^- - \frac{1}{\ell^2} \bar{\varphi}^+ \gamma^0
\chi - \frac{1}{\ell^2} \bar{\zeta} \gamma^0 \xi^+ \,,
\notag \\
\delta \psi^{+} &=&d\epsilon ^{+}+\frac{1}{2}\omega \gamma _{0}\epsilon
^{+} - \frac{1}{2} y_1 \gamma_0 \epsilon^+ +\frac{1}{2 \ell^2}\tau \gamma
_{0}\varphi ^{+} +\frac{1}{2 \ell^2}k \gamma _{0}\epsilon ^{+} - \frac{1}{2
\ell^2} u_1 \gamma_0 \varphi^+ - \frac{1}{2 \ell^2} b_1 \gamma_0 \epsilon^+
\,,  \notag \\
\delta \psi^{-} &=&d\epsilon ^{-}+\frac{1}{2}\omega \gamma _{0}\epsilon
^{-}+\frac{1}{2}\omega ^{a}\gamma _{a}\epsilon ^{+} + \frac{1}{2} y_1
\gamma_0 \epsilon^- +\frac{1}{2 \ell^2}\tau \gamma _{0}\varphi ^{-} +\frac{1%
}{2\ell^2}e ^{a}\gamma _{a}\varphi ^{+} +\frac{1}{2\ell^2}k ^{a}\gamma
_{a}\epsilon ^{+}  \notag \\
&& +\frac{1}{2 \ell^2}k \gamma _{0}\epsilon ^{-} + \frac{1}{2 \ell^2} u_1
\gamma_0 \varphi^- + \frac{1}{2 \ell^2} b_1 \gamma_0 \epsilon^- \,,  \notag
\\
\delta \xi ^{+} &=&d\varphi ^{+}+\frac{1}{2}\omega \gamma _{0}\varphi
^{+}+\frac{1}{2}\tau \gamma _{0}\epsilon ^{+} - \frac{1}{2} y_1 \gamma_0
\varphi^+ - \frac{1}{2} u_1 \gamma_0 \epsilon^+ + \frac{1}{2 \ell^2} k
\gamma_0 \varphi^+ - \frac{1}{2 \ell^2} b_1 \gamma_0 \varphi^+ \,,  \notag \\
\delta \xi ^{-} &=&d\varphi ^{-}+\frac{1}{2}\omega \gamma _{0}\varphi
^{-} +\frac{1}{2}\tau \gamma_0 \epsilon^- + \frac{1}{2} e^a \gamma_a
\epsilon^+ +\frac{1}{2}\omega ^{a}\gamma _{a}\varphi ^{+}+\frac{1}{2} y_1
\gamma _{0}\varphi^{-}+ \frac{1}{2} u_1 \gamma_0 \epsilon^-  \notag \\
&& + \frac{1}{2 \ell^2} k^a \gamma_a \varphi^+ + \frac{1}{2 \ell^2} k
\gamma_0 \varphi^- + \frac{1}{2 \ell^2} b_1 \gamma_0 \varphi^- \,,  \notag \\
\delta \rho &=&d\eta +\frac{1}{2}\omega \gamma _{0}\eta +\frac{1}{2}\omega
^{a}\gamma _{a}\epsilon ^{-}+\frac{1}{2}s\gamma _{0}\epsilon ^{+} - \frac{1}{%
2} y_2 \gamma_0 \epsilon^+ - \frac{1}{2} y_1 \gamma_0 \eta + \frac{1}{2
\ell^2} e^a \gamma_a \varphi^-  \notag \\
&& + \frac{1}{2 \ell^2} k^a \gamma_a \epsilon^- + \frac{1}{2 \ell^2} m
\gamma_0 \varphi^+ + \frac{1}{2 \ell^2} \tau \gamma_0 \zeta + \frac{1}{2
\ell^2} t \gamma_0 \epsilon^+ + \frac{1}{2 \ell^2} k \gamma_0 \eta  \notag \\
&& - \frac{1}{2 \ell^2} u_1 \gamma_0 \zeta - \frac{1}{2 \ell^2} u_2 \gamma_0
\varphi^+ - \frac{1}{2 \ell^2} b_1 \gamma_0 \eta - \frac{1}{2 \ell^2} b_2
\gamma_0 \epsilon^+ \,,  \notag \\
\delta \chi &=&d\zeta + \frac{1}{2}\omega \gamma _{0}\zeta + \frac{1}{2}%
\omega ^{a}\gamma _{a}\varphi ^{-}+\frac{1}{2}e^{a}\gamma _{a}\epsilon ^{-}+%
\frac{1}{2}\tau \gamma _{0}\eta +\frac{1}{2}s\gamma _{0}\varphi ^{+}+\frac{1%
}{2}m\gamma _{0}\epsilon ^{+}  \notag \\
&& - \frac{1}{2} y_2 \gamma_0 \varphi^+ - \frac{1}{2} y_1 \gamma_0 \zeta -
\frac{1}{2} u_2 \gamma_0 \epsilon^+ - \frac{1}{2} u_1 \gamma_0 \eta + \frac{1%
}{2 \ell^2} k^a \gamma_a \varphi^- + \frac{1}{2 \ell^2} t \gamma_0 \varphi^+
+ \frac{1}{2 \ell^2} k \gamma_0 \zeta  \notag \\
&& - \frac{1}{2 \ell^2} b_1 \gamma_0 \zeta - \frac{1}{2 \ell^2} b_2 \gamma_0
\varphi^+ \,,  \label{susytr}
\end{eqnarray}
where $\epsilon ^{\pm }$, $\varphi ^{\pm }$, $\eta $, and $\zeta $ are the
fermionic gauge parameters related to $\tilde{Q}^{\pm }$, $\tilde{\Sigma}
^{\pm }$, $\tilde{R}$, and $\tilde{W}$, respectively. Let us observe
that by taking the flat limit $\ell \rightarrow \infty$ of 
one recovers the supersymmetry transformations under which the curvatures
associated with the MEB superalgebra transform in a covariant way (see \cite{Concha:2019mxx}).


\bibliographystyle{fullsort}
 
\bibliography{Non_relativistic_super_AdS-Lorentz_V5}

\end{document}